\documentclass{llncs}
\usepackage{pifont}
\usepackage{llncsdoc}
\usepackage{bbm}
\usepackage{amsmath}
\usepackage{amsfonts}
\usepackage{mathrsfs}
\usepackage{amssymb}
\usepackage{graphicx}
\usepackage{epsfig}
\usepackage{verbatim}
\usepackage[vlined,ruled]{algorithm2e}
\usepackage{sidecap}

\usepackage{float}
\floatstyle{ruled}
\restylefloat{figure}

\newtheorem{df1}{Definition}
\newtheorem{th1}{Theorem}
\newtheorem{le1}{Lemma}

\newtheorem{pr1}{Proposition}

\newcommand{\hbem}[1]{\hbox{\emph{#1}}}
\newcommand{\2}{\mid}
\newcommand{\rec}[1]{[\mathrm{rec}\,#1]}

\newcommand{\lrceil}[1]{\lceil #1 \rceil}
\newcommand{\ldot}{.\;}
\newcommand{\tsum}{\sqcup}
\newcommand{\etal}{\emph{et al.~}}

\begin{document}

\title{Session Communication and Integration}
\author{Guoxin Su\inst{1,3}\and Mingsheng Ying\inst{1,2}\and Chengqi Zhang\inst{1}}
\institute{
Centre for Quantum Computation $\&$ Intelligent Systems\\  University of Technology, Sydney\and
State Key Laboratory of Intelligent Technology $\&$ Systems\\
Tsinghua University\\
\and\email{guoxin@it.uts.edu.au}
}

\maketitle

\begin{abstract}
The scenario-based specification of a large distributed system is usually naturally decomposed into various modules. The integration of specification modules contrasts to the parallel composition of program components, and includes various ways such as scenario concatenation, choice, and nesting. The recent development of multiparty session types for process calculi provides useful techniques to accommodate the protocol modularisation, by encoding fragments of communication protocols in the usage of private channels for a class of agents. In this paper, we extend forgoing session type theories by enhancing the session integration mechanism. More specifically, we propose a novel synchronous multiparty session type theory, in which sessions are separated into the communicating and integrating levels. Communicating sessions record the message-based communications between multiple agents, whilst integrating sessions describe the integration of communicating ones. A two-level session type system is developed for $\pi$-calculus with syntactic primitives for session establishment, and several key properties of the type system are studied. Applying the theory to system description, we show that a channel safety property and a session conformance property can be analysed. Also, to improve the utility of the theory, a process slicing method is used to help identify the violated sessions in the type checking.
\end{abstract}


\section{Introduction}

The description of service accesses in protocols has been long considered as a way to improve the interoperability of program components in a complex computing system, and this is the case for various architecture description languages (e.g.~Darwin \cite{mdek95}, Wright \cite{ag97}, and PADL \cite{bcd02}) and component-based platforms (e.g.~Coyote \cite{bhsc98} and Appia \cite{mpr01}). Formal validation methods including model checking and static checking are employed to aid the detection of composition mistakes such as deadlocks and race conditions. For large distributed systems, the specification is usually modularised. Studies on session types \cite{hvk98,gh99} for process calculi in the dialect of $\pi$-calculus \cite{sw01}, especially the recent development of multiparty session types \cite{hyc08}, provide useful techniques to accommodate the protocol modularisation. Informally, a session is a unit of message-based communications with a specific purpose. The suitability of session type theories for describing distributed computing includes two aspects:
\begin{itemize}
  \item Session type theories provide a global descriptive method for protocols, facilitating the protocol design and verification;
  \item To handle protocol modularisation, type systems in session type theories project fragments of the protocols on the usage of private channels for intended classes of participants.
\end{itemize}

However, the origin of session type theories assumes the interleaving situation of different independent behavioural threads \cite{hmbcy11}, but in real-life distributed computing, there are often meaningful interplays between a number of sessions. As an example, the following business protocol consists of four sessions between five agents. A broker and two buyers are in the auction session $\hbem{Auction}$ (where $\hbem{Auction}$ is seen as a global description of the auction protocol). $\hbem{Auction}$ is followed by a transaction session between the auction winner, the broker, and the seller. Two alternative transaction protocols are given for the winner to choose: $\hbem{DTransaction}$ is a direct protocol, in which the winner directly transfers money to the seller; $\hbem{STransaction}$ is a secure protocol, in which the money is transferred via the broker and an extra (sub-)session $\hbem{EPay}$ money transaction between the winner, the bank, and the broker is involved. Therefore, the whole business protocol is the integration of four sessions in the intended ways. It is indeed possible to view the protocol as inseparable, but so-doing violates both the natural understanding of the protocol and the gradual procedure of requirement specification.

To improve the session integration mechanism for session type theories, we argue for the merits of separating sessions into two levels. In the present paper, we propose a theory of two-level synchronous multiparty session types, in which communicating sessions specify the end-point communications of multiple components, whilst integrating sessions describe the gluing of communicating sessions by concatenation, choice, interleaving composition, and nesting composition. Compared with the existing studies in this subject, e.g.~\cite{chy07,hyc08,ydbh10,dy11,dh12}, in addition to the separation of session communication and integration, the novelty of our work includes the following aspects. First, we view sessions as a behavioural rather than data-structural approximation of processes. Besides for statically typing processes in a variant of $\pi$-calculus, session types are also executable and equipped with intuitive operational semantics. In the forgoing session type theories, sessions as the specification leave out data structures required in the implementation, but based on the operational semantics of sessions, we investigate the behavioural relation between processes and sessions. We demonstrate that, in spite of session modularisation and integration, behaviours of processes, if typed properly, conform to their session specification. Second, the most recent work on session types witnesses a trend to introduce more expressive session constructs and, correspondingly, more syntactic primitives in the underlying calculi, but the ramification of communicating and integrating sessions in our work does not complicate the syntax of the process calculus. Lastly, to improve the utility of the theory, we use a process slicing method to help identify the violated sessions in the type checking. The method decomposes a process into parts with respect to sessions in its session specification and compares each part with the role projected from a corresponding session.


The organisation of the remainder of the paper is as follows. In the next section, we present a process calculus with actions for multiparty session establishment. In Sect.~\ref{sec:sessiontype}, we define the syntax and semantics of communicating and integrating sessions, together with methods to project sessions into roles for processes. In Sect.~\ref{sec:typediscipline}, we develop a two-level session type system and study several key properties of the type system. In Sect.~\ref{sec:analysis}, we apply the session type theory to system description and analyse a channel safety property and a behavioural property of session conformance. In Sect.~\ref{sec:slicing}, we use a process slicing method to facilitate the identification of violated sessions in type checking. In Sect.~\ref{sec:relatedwork}, we discuss the related work to this paper. Finally, we conclude the paper by outlining the future work. 
More examples and proof details of the theorems are in the Appendix.

\section{The Calculus}\label{sec:calculus}

\begin{figure}[t]\vspace{-0.8em}
\begin{align*}
     P ::= & ~\pi. P & \hbox{prefixing} & \qquad\qquad |~~ X & \hbox{variable}\\
     |~ & ~ (\nu a) P & \hbox{hiding} & \qquad\qquad |~~ \rec{X}P & \hbox{recursion}\\
     |~ & ~ r:P & \hbox{labelling} & \qquad\qquad |~~ P+P & \hbox{choice}\\
     |~ & ~ \mathbf{0} & \hbox{inaction} & \qquad\qquad |~~ P\2 P & \hbox{parallel}  \\
    \alpha ::= &~ \pi & \hbox{action} & \qquad\qquad |~~ \tau & \hbox{silence} \\
    \pi ::= & ~ \bar{a}_{[2..n]}(\tilde{c})& \hbox{invitation} & \qquad\qquad |~~ a?v & \hbox{receiving}\\
     |~ & ~ a_{[k]}(\tilde{c})  & \hbox{acceptance} & \qquad\qquad |~~ a!v & \hbox{sending}
\end{align*}
\caption{Syntax of the calculus\label{syntax}}\vspace{-1em}
\end{figure}

This section defines a variant of $\pi$-calculus. In the next two sessions, a type discipline based on two-level session types is developed for the calculus. Compared with the existing session type literature, the syntax of our calculus is abstract and close to the original presentation of $\pi$-calculus. Our intention is to minimise the side techniques (we return to this point in Sect.~\ref{sec:relatedwork}).

The basic sets are a set of \emph{channels} ($a,b,c,c'$), a set of \emph{messages} or \emph{message types} ($u,v,v'$), and a set of \emph{participant names} ($p,q,r,1,2, \,\hbox{ect.}$). The syntax of \emph{processes} and \emph{actions} is given in Figure \ref{syntax}. Sessions, which are informally understood as units of interactions, are established by shared channels. The key syntactic primitives for channel establishment are of the forms $\bar{a}_{[2..n]}(\tilde{c})$ and $a_{[k]}(\tilde{c})$, which are due to \cite{hyc08}. These two prefixes are called \emph{session actions} and $a$ is called a \emph{session channel}. $\bar{a}_{[2..n]}(\tilde{c})$ invites participants $2$ to $n$ to join in a session whose communicating channels are $\tilde{c}$, whilst $a_{[k]}(\tilde{c})$ accepts a session invitation. By the operational semantics, when the actions $\bar{a}_{[2..n]}(\tilde{c})$ and $a_{[k]}(\tilde{c})$ (for each $2\leq k\leq n$) are triggered synchronously, a session is established via the session channel $a$ and a sequence of fresh communicating channels $\tilde{c}$ are generated (unlike \cite{hyc08}, in which the message transport is asynchronous). In $r:P$, $r$ labels $P$ and is seen as the name of $P$. In some literature, it is also called the location of $P$ \cite{hen07}. Other syntactic primitives and constructions are standard and from $\pi$-calculus.

Binders are $a$ in $(\nu a) P$, $\tilde{c}$ in $\bar{a}_{[2..n]} (\tilde{c}).P$ or $a_{[k]}(\tilde{c}).P$, and $X$ in $\rec{X}P$. Substitution of channels are standard. In particular, $(\rec{X}P)\{a/b\}=\rec{X}(P\{a$ $/b\})$. $(\nu \tilde{a})P$ stands for $(\nu a_1)\ldots (\nu a_n)P$ where $\tilde{a}=a_1,\ldots, a_n$. The left-\-associative law is adopted when presenting multiple $\2$ or $+$. We assume the bound name convention for processes. Let $\mathrm{fc}(P)$ and $\mathrm{fc}(\alpha)$ denote the set of free channels in $P$ and $\alpha$, respectively. $\mathrm{fv}(P)$ is the set of free process variables, and $\mathrm{act}(P)$ the set of prefixes in $P$.
Supposing $\tilde{a}=\mathrm{fc}(P)$ and $|\tilde{c}|=|\tilde{a}|$, $P\langle \tilde{c}\rangle$ refers $P\{ \tilde{c}/\tilde{a}\}$.

The structural congruence $\equiv$ is the smallest congruent relation on processes that includes the equations in Figure \ref{cong}. $P=_\alpha Q$ means that $P$ and $Q$ are variants of \emph{alpha-conversion}. Note that we leave out the equi-recursive equation (e.g.~$\rec{X}P\equiv P\{\rec{X}P/X\}$) in the structural laws (it is called recursion-free or replication-free structural congruence in some literature \cite{eg04}). Consequently, we have the decidability of structural congruence.
\begin{le1}\label{le:process-equivalence}
For any given $P,Q$, it is decidable if $P\equiv Q$.

\end{le1}
\begin{figure}[t]\vspace{-1em}
\begin{gather*}
    P\2 Q\equiv Q\2 P\qquad P\2 \mathbf{0}\equiv P \qquad
    (P\2 Q)\2 R\equiv P\2 (Q\2 R)\\
    P+Q\equiv Q+P\qquad  P+\mathbf{0}\equiv P \qquad
  (P+Q)+R\equiv P+(Q+R) \\
     (\nu a) \mathbf{0}\equiv \mathbf{0} \qquad (\nu a)(\nu b)P\equiv (\nu b)(\nu a) P\qquad
    (\nu a)P\2 Q \equiv (\nu a )(P\2 Q) \;\hbox{ if } a\notin\mathrm{fc}(Q) \\
     \rec{X}\mathbf{0}\equiv\mathbf{0} \qquad \rec{X}R\equiv R \;\hbox{ if } X\not\in \mathrm{fv}(R)\qquad
    P\equiv Q \;\hbox { if } P=_\alpha Q\\
     l':l:P\equiv l':P   \qquad  l:P\2 l:Q\equiv l:(P\2 Q) \qquad  l:(\nu a) P\equiv (\nu a)l:P
\end{gather*}\vspace{-1.5em}
\caption{Structural congruence}\label{cong}
\end{figure}

If $X\in \mathrm{fv}(P)$, we let $P^{[X]}$ be $\rec{X} P$; otherwise, $P^{[X]}$  is $P$. $P\sqsubseteq Q$ means $P+R\equiv Q$ for some $R$, and $P\sqsubset Q$ means $P+R\equiv Q$ for some $R\not\equiv \mathbf{0}$. $P\sqcup Q$ is defined as follows: if $Q\sqsubseteq P$ then $P\sqcup Q= P$; if $P\sqsubset Q$ then $P\sqcup Q= Q$; otherwise, $P \sqcup Q= P+Q$.

The operational semantics are given through a labelled transition system defined by rules in Figure \ref{tran}. In the session type literature, the semantics of the process calculus is defined as a reduction system instead of a labelled transition one. The advantage of the former over the latter is a simpler presentation. But because one of our purposes in the present paper is to study the behavioural relation between processes and their session types, the standard operational semantics are more suitable to this end. The rules [\textsc{Inv}], [\textsc{Acc}], and [\textsc{Sess}] handle the session establishment, and their intuitive meanings have been explained. [\textsc{Lab}] is for process labelling.  The rest of the semantic rules are standard.

Let $\mathrm{proc}(P)=\{Q~|~P\overset{\tilde{\alpha}}{\longrightarrow}Q\}$.
$P$ \emph{stimulates} $Q$, denoted $P\succ Q$, if there is a relation $\mathcal{S}\subseteq \mathrm{proc}(P)\times \mathrm{proc}(Q)$ such that
if $\langle P,Q\rangle \in \mathcal{S}$ and $Q\overset{\alpha}{\longrightarrow}Q'$ then there is $P'$ such that $P\overset{\alpha}{\longrightarrow} P'$ and $\langle P',Q'\rangle\in \mathcal{S}$. We use $P \overset{\alpha}{\longrightarrow}_\succ Q$ to mean that there is $R$ such that $P \overset{\alpha}{\longrightarrow} R$ and $R\succ Q$. We say $P$ is \emph{deterministic} (up to structural congruence) if for each $Q\in \mathrm{proc}(P)$, $Q\overset{\alpha}{\longrightarrow}R$ and $Q\overset{\alpha}{\longrightarrow}R'$ entail $R\equiv R'$.

\begin{figure}[t]\vspace{-0.8em}
\begin{gather*}
      \hbox{[\textsc{Sed}]}~~ a! v.P\overset{a! v}{\longrightarrow} P \qquad \hbox{[\textsc{Rcv}]}~~ a? v.P\overset{a? v}{\longrightarrow} P \\
       \hbox{[\textsc{Inv}]}~~ \bar{a}_{[2..n]}(\tilde{c}).P\overset{\bar{a}_{[2..n]}(\tilde{c})}{\longrightarrow} P    \qquad
      \hbox{[\textsc{Acc}]} ~~a_{[k]}(\tilde{c}).P\overset{a_{[k]}(\tilde{c}')}{\longrightarrow} P\{\tilde{c}'/\tilde{c}\} \\
    \hbox{[\textsc{Par}]}~~\dfrac{P\overset{\alpha}{\longrightarrow} P'}{P\2 Q \overset{\alpha}{\longrightarrow}P'\2 Q} \qquad \hbox{[\textsc{Sum}]}~~
    \dfrac{P\overset{\alpha}{\longrightarrow} P'}{P+ Q \overset{\alpha}{\longrightarrow}P'}\\
    \hbox{[\textsc{Com}]}~~\dfrac{P\overset{a!v}{\longrightarrow} P' \qquad Q\overset{a? v}{\longrightarrow} Q' }{P\2 Q\overset{\tau}{\longrightarrow}P'\2 Q}\qquad  \hbox{[\textsc{Hid}]}~~ \dfrac{P\overset{\alpha}{\longrightarrow} P'\quad \mathrm{fc}(\alpha)\neq a}{(\nu a)P\overset{\alpha}{\longrightarrow} (\nu a) P'}\\
    \hbox{[\textsc{Sess}]}~~\dfrac{P_1\overset{\bar{a}_{[2..n]}(\tilde{c})}{\longrightarrow} P_1' \quad P_i\overset{a_{[i]}(\tilde{c})}{\longrightarrow} P_i'~( \forall\, 2\leq i\leq n )}
    {P_1\2\ldots \2 P_{n}\overset{\tau}{\longrightarrow} (\nu \tilde{c})( P_1'\2\ldots\2 P_{n}')} \qquad \hbox{[\textsc{Lab}]} ~~\dfrac{P\overset{\alpha}{\longrightarrow} P'} { l:P\overset{\alpha}{\longrightarrow} l:P'}\\
    \hbox{[\textsc{Rec}]}~~ \dfrac{P\overset{\alpha}{\longrightarrow} P'}{\rec{X}P\overset{\alpha}{\longrightarrow} P'\{\rec{X}P/X\}} \qquad \hbox{[\textsc{Eqv}]}~~
     \dfrac{P\equiv Q\quad Q\overset{\alpha}{\longrightarrow}Q'\quad Q'\equiv P'}{P\overset{\alpha}{\longrightarrow}P'}
\end{gather*}\vspace{-0.6em}
\caption{Operational semantics}\label{tran}
\end{figure}

\paragraph{\textbf{Examples of agent behaviours}}
We provide a detailed, but informal description of the interactions between the five agents in the example from Introduction, and then formulate their individual behaviours in the calculus. In the upmost level, the whole business protocol is divided into two stages. The first stage is for the auction session and the second one includes two alternative transaction sessions and a possible nested sub-session.

At the auction stage, the broker initiates the auction session with two buyers, i.e.~$\mathrm{buyer}_1$ and $\mathrm{buyer}_2$. For simplicity, we assume that the buyers already know the base price of the auctioned item. After the auction is initiated, $\mathrm{buyer}_1$ (resp.~$\mathrm{buyer}_2$) sends its bid to the broker and the protocol reaches a recursive state. In the recursive state, the broker sends a new quote to the other buyer and the protocol proceeds in the following two alternative branches. (a) If the other buyer does not bid (after some amount of time), then the broker issues an invoice to $\mathrm{buyer}_1$ (resp.~$\mathrm{buyer}_2$), finishing the auction. (b) If the other buyer bids, then the broker forwards the latest quote to $\mathrm{buyer}_1$ (resp.~$\mathrm{buyer}_2$) and, again, the protocol has two sub-branches: (b1) if $\mathrm{buyer}_1$ (resp.~$\mathrm{buyer}_2$) continues to bid, then the protocol returns to the recursive state; (b2) otherwise, the broker issues $\mathrm{buyer}_2$ (resp.~$\mathrm{buyer}_1$) an invoice to finish the auction.

At the transaction stage, the buyer that won the auction initiates one of the following two transaction options. (a) If the direct transaction is chosen, then the broker forwards the price to the seller, and the buyer makes the payment to the seller and receives the ordering information. (b) If the secure transaction is chosen, an extra bank transfer session is involved. The buyer authorises the bank to transfer an intended amount of money to the broker. The broker holds the money but informs the seller that the pre-payment is ready, and seller sends the ordering information to the buyer. After receiving the item, the buyer sends a confirmation message to the broker and the broker finalises the deal by transferring the pre-payment to the seller.

The formal description of the broker's behaviour is given by the following process.
\begin{align*}
    P_{\mathrm{broker}} \overset{\mathrm{def}}{=} ~& \overline{\mathrm{auc}}_{[2..3]}(a_{1,2},a_{1,3})\ldot\sum_{i\in \{1,2\}}(a_{1,i+1}?\mathrm{bid}\ldot\rec{X}(a_{1,4-i}!\mathrm{quote}\ldot \\
    & \qquad (a_{1,i+1}!\mathrm{invoice}\ldot P_{\mathrm{broker}}^i +a_{1,4-i}?\mathrm{bid}\ldot a_{1,i+1}!\mathrm{quote}\ldot\\
    & \qquad\quad (a_{1,i+1}?\mathrm{bid}\ldot X+a_{1,4-i}!\mathrm{invoice}\ldot P_{\mathrm{broker}}^{3-i})))) \\
    P_{\mathrm{broker}}^i \overset{\mathrm{def}}{=} ~& \mathrm{dTran}_{[2]}^i(b_{1,2},b_{2,3})\ldot b_{2,3}!\mathrm{price}\ldot\mathbf{0}+\mathrm{sTran}_{[2]}^i(c_{1,2},c_{1,3},c_{2,3})\ldot \\
& \qquad  \mathrm{epay}_{[2]}^i(d_{1,3}, d_{2,3})\ldot d_{1,3}?\mathrm{transfer}\ldot c_{2,3}!\mathrm{prepaid}\ldot \\
& \qquad \quad  c_{1,2}?\mathrm{confirm}\ldot c_{2,3}!\mathrm{payment}\ldot \mathbf{0}
\end{align*}

As explained before, the session is established via shared channels, i.e.~session channels, one of which is $\mathrm{auc}$.
The prefix $\overline{\mathrm{auc}}_{[2..3]}(a_{1,2},a_{1,3})$ initiates a session with another two participants (three in total), which, in this case, are the two buyers. $\mathrm{dTran}_{[2]}^i(b_{1,2},b_{2,3})$, $\mathrm{sTran}_{[2]}^i(c_{1,2},c_{1,3},c_{2,3})$, and $\mathrm{epay}_{[2]}^i(d_{1,3},d_{2,3})$ ($i\in \{1,2\}$) are for accepting session establishment. Other prefixes are ordinary prefixes in $\pi$-calculus. We use $a_{i,j}$ to denote the channel used by the $i$th and $j$th agents in the session. For example, $a_{1,2}$ is the communicating channel between the first and second participants, which, in this case, are the broker and $\mathrm{buyer}_1$. The recursive structure in $P_{\mathrm{broker}}$ corresponds to the recursive state of the auction protocol informally described above.

The behaviours of the two buyers in the two stages of the protocol follow. Let $j\in \{1,2\}$.
\begin{align*}
    P_{\mathrm{buyer}_j}\overset{\mathrm{def}}{=} ~ &  \mathrm{auc}_{[j+1]}(a_{1,2},a_{1,3})\ldot a_{1,j+1}!\mathrm{bid}\ldot\rec{X_1}(a_{1,j+1}?\mathrm{quote}\ldot a_{1,j+1}!\mathrm{bid}\ldot X \\
     & \quad + a_{1,j+1}?\mathrm{invoice}\ldot P_{\mathrm{buyer}_i}')+ a_{1,j+1}?\mathrm{quote}\ldot\rec{X_2} \\
     & \quad \quad  (a_{1,j+1}!\mathrm{bid}\ldot (a_{1,j+1}?\mathrm{invoice}\ldot P_{\mathrm{buyer}_j}'+a_{1,j+1}?\mathrm{quote}\ldot X_2)) 
\\
    P'_{\mathrm{buyer}_j}\overset{\mathrm{def}}{=} ~& \overline{\mathrm{dTran}}_{[2..3]}^j(b_{1,3},b_{2,3})\ldot b_{1,3}!\mathrm{payment}\ldot b_{1,3}?\mathrm{order}\ldot \mathbf{0} \; + \\
     & \quad  \overline{\mathrm{sTran}}_{[2..3]}^j(c_{1,2},c_{1,3},c_{2,3})\ldot \overline{\mathrm{epay}}_{[2..3]}(d_{1,3},d_{2,3})\ldot d_{1,3}!\mathrm{amount}\ldot \\
     & \quad \quad  d_{2,3}?\mathrm{transfer}\ldot c_{1,3}?\mathrm{order}\ldot \mathrm{c}_{1,2}!\mathrm{confirm}\ldot \mathbf{0}
\end{align*}

The seller and the bank only take part in the second part of the protocol, and their behaivours follow.
\begin{align*}
    P_{\mathrm{seller}} \overset{\mathrm{def}}{=} & \sum_{i\in\{1,2\}}( \mathrm{dTran}_{[3]}^i(b_{1,3},b_{2,3})\ldot b_{2,3}?\mathrm{price}\ldot b_{1,3}?\mathrm{payment}\ldot b_{1,3}!\mathrm{order}\ldot \mathbf{0}\; \\
    & \,+\,\mathrm{sTran}_{[3]}^i(c_{1,2},c_{1,3},c_{2,3})\ldot c_{2,3}?\mathrm{prepaid}\ldot c_{1,3}!\mathrm{order}\ldot c_{1,3}!\mathrm{payment}\ldot \mathbf{0} )\\
    P_{\mathrm{bank}}\overset{\mathrm{def}}{=} &\, \mathrm{epay}_{[3]}(d_{1,3},d_{2,3})\ldot d_{1,3}?\mathrm{amount}\ldot d_{2,3}!\mathrm{transfer}\ldot \mathbf{0}
\end{align*}

The interactions of five processes, i.e.~$P_{\mathrm{broker}}$, $P_{\mathrm{buyer}_1}$, $P_{\mathrm{buyer}_2}$, $P_{\mathrm{seller}}$, and $P_{\mathrm{bank}}$ according the operational semantics should follow the presented scenario.

\section{Two-level Session Types}\label{sec:sessiontype}

\subsection{Syntax and Semantics}

\begin{figure}[t]\vspace{-1em}
\begin{align*}
   S,T :: = &~ \langle p,q:v\rangle \rightarrow S & \hbox{communication}& \qquad\qquad |~~ \mathtt{end} & \hbox{termination} \\
        |~&~ \langle \tilde{p}:S\rangle\{T\} & \hbox{establishment}& \qquad\qquad |~~ S;T & \hbox{concatenation}\\
        |~&~ \mathbf{t} & \hbox{type variable}& \qquad\qquad |~~ S\oplus T & \hbox{union}\\
        |~&~ \mu \mathbf{t}.S & \hbox{recursion}& \qquad\qquad |~~ S\otimes T & \hbox{product}
\end{align*}\vspace{-1.5em}
\caption{Abstract syntax of sessions}\label{ses:type}
\end{figure}

We provide the general syntax of \emph{session types} or \emph{sessions}, and then define the two kinds of sessions studied in the present paper, i.e.~communicating and integrating sessions.

The general syntax is provided in Figure \ref{ses:type}. $\langle p,q:v\rangle \rightarrow S$ is a session of \emph{communication} form, meaning that, after the agent $p$ sends the message (type) $v$ to the agent $q$, the session proceeds as $S$. $\langle \tilde{p}:S\rangle\{T\}$ is a session of \emph{establishment} form, meaning that the agents $\tilde{p}$ establish a session $S$ which nests $T$. The first item in the sequence $\tilde{p}$ refers to the participant that initiates $S$.
We call $\langle p,q:v\rangle$ and $\langle \tilde{p}:S\rangle$ \emph{event prefixes}. $S;T$ is the \emph{concatenation} of $S$ and $T$, $S\oplus T$ is their \emph{union}, and $S\otimes T$ their \emph{product}. $\mathbf{t}$ is a type variable and $\mu\mathbf{t}. S$ is a recursive type and binds $\mathbf{t}$ in $S$ in the standard way. A session is \emph{close} if occurrences of all variables in it are bound. $\mathtt{end}$ is a terminated session.

If $T$ is is contained in (the presentation of) $S$, we call $T$ a sub-session of $S$. $\mathrm{pid}(S)$ is the set of participant names in $S$.
$S$ is \emph{well-formed}, if (1) for each sub-session $\langle p,q:v\rangle \rightarrow T$ of $S$, $p\neq q$, (2) for each sub-session $\langle \tilde{p}:T\rangle\{T'\}$ of $S$, $\tilde{p}$ is distinct and $|\tilde{p}|= |\mathrm{pid}(T)|$, and (3) for each sub-session $T;T'$ of $S$, $T$ does not contain $\otimes$. Hence, well-formedness of sessions rules out self interactions such as $\langle p,p:v\rangle$ and multiple participation of sessions such as $\langle p,p:S\rangle$. We also require the left session of a concatenated session to be single-threaded.
For each $S$, we define a set $\mathrm{opid}(S)$ as follows:
\begin{itemize}
  \item $\mathrm{opid}(\mathbf{t})=\mathrm{opid}(\mathtt{end})=\varnothing$; $\mathrm{opid}(\mu\mathbf{t}.S)=\mathrm{opid}(S)$;
  \item $\mathrm{opid}( \langle p,q:v\rangle \rightarrow S)=\{ \{ p,q\}\}$; $\mathrm{opid}( \langle \tilde{p}:S\rangle\{T\})=\{  \tilde{p}\}$;
  \item $\mathrm{opid}(S\oplus T)=\mathrm{opid}(S\otimes T)=\mathrm{opid}(S)\cup \mathrm{opid}(T)$;
  \item \begin{itemize}
          \item if $\mathrm{opid}(S)\neq \varnothing$, then $\mathrm{opid}(S; T)=\mathrm{opid}(S)$; and
          \item if $\mathrm{opid}(S)=\varnothing$, then $\mathrm{opid}(S; T)=\mathrm{opid}(T)$.
        \end{itemize}
\end{itemize}
Then, \emph{race-free} sessions are recursively defined as follows. (1) $\mathtt{end}$ and $\mathrm{opid}(\mathbf{t})$ are race-free; (2) if $S$ is race-free, then so is $\mu\mathbf{t}.S$; (3) if $S$ is race-free and $\{p,q\}\cap H\neq \varnothing$ for each $H\in\mathrm{opid}(S)$, then $\langle p,q:v\rangle\rightarrow S$ is race-free for any $v$; (4) if $S,H$ are race-free, then $\langle \tilde{p}:S\rangle \{T\}$ is race-free; (5) if $S, T$ are race-free and $H\cap H'\neq \varnothing$ for each $H\in \mathrm{opid}(S),H'\in \mathrm{opid}(T)$, then $S\oplus T$ is race-free; (6) if $S, T$ are race-free then $S\otimes T$ is race-free; (7) if $T$ is race-free and $\mathrm{pid}(S)=\varnothing$, then $S;T$ is race-free; and (8) if $S,T$ are race-free, $\mathrm{pid}(S)\neq\varnothing$, and $\mathrm{pid}(S')\cap H\neq\varnothing$ for each $H\in \mathrm{opid}(T)$ and each subsession $S'$ of $S$, then $S;T$ is race-free. Let $\tilde{p}$ be a distinct sequence and $|\tilde{p}|=\mathrm{pid}(S)$. We use $S\langle \tilde{p}\rangle$ to denote the simultaneous substitution of $\tilde{p}$ for $\mathrm{pid}(S)$ in $S$.

In the following, we restrict the general syntax of sessions and define two special kinds of sessions studied in the present paper.
\begin{df1}[Communicating sessions]\label{df:comsession} The syntax of communicating sessions ($B,B'$) contains rules in Figure \ref{ses:type} except the establishment (i.e.~$\langle \tilde{p}:S\rangle\{T\}$). \end{df1}

For simplicity, for a given communicating session $B$, we let $\mathrm{pid}(B)$ be a sequence of consecutive integral numbers from $1$. \emph{Auction}, \emph{DTransaction}, \emph{STransaction}, and \emph{EPay} in Sect.~\ref{ses:sessionexample} below are communicating sessions. However, when writing $B\langle \tilde{p}\rangle$ , $\tilde{p}$ is not necessarily a sequence of consecutive integrals.

\begin{df1}[Integrating sessions]\label{df:intsession}
The syntactic rule of restricted establishment is a restricted one of the establishment in Figure \ref{ses:type}: $\langle \tilde{p}: B\rangle \{T\}$ where $T$ is a general session and $B$ is a communicating session.
The syntax of integrating sessions ($A,A'$) contains the restricted establishment and constructions in Figure \ref{ses:type} except the establishment and communication (i.e.~$\langle p,q:v\rangle \rightarrow S$ in Figure \ref{ses:type}).
\end{df1}

\begin{figure}[t]\vspace{-1em}
\begin{gather*}
    (S\oplus S')\oplus S''\equiv S\oplus(S'\oplus S'')\qquad
    S\oplus S' \equiv S'\oplus S \qquad S\oplus \mathtt{end} \equiv S \\
    (S\otimes S')\otimes S''\equiv S\otimes (S'\otimes S'')\qquad
    S\otimes S' \equiv S'\otimes S \qquad  S\otimes \mathtt{end} \equiv S\\
    (S;S');S''\equiv S;(S';S'') \qquad S ;\mathtt{end}\equiv S \qquad  \mathtt{end}; S \equiv S
      \\
  \mu \mathbf{t}. S \equiv S ~\hbox{ if } \mathbf{t}\not\in \mathrm{fv}(S) \qquad  S\equiv T ~\hbox{ if } S=_\alpha T
\end{gather*}\vspace{-1.5em}\caption{Session structural congruence}\label{ses:stru}
\end{figure}

\emph{Proto} in Sect.~\ref{ses:sessionexample} below is an integrating session. We use $\langle \tilde{p}:S\rangle$ to abbreviate $\langle \tilde{p}:S\rangle\{\mathtt{end}\}$.
In the sequel, we assume that \emph{the communicating and integrating sessions under consideration are well-formed and race-free.} Well-formedness requirement is due to syntactic legitimacy. We additionally require sessions to be race-free, because otherwise their process-level counterparts (obtained by the projection method described below) may contain race-conditions. Due to space limitations, we leave the detailed discussion for future work.

The structural congruence of sessions is the smallest congruent relation containing the equations presented in Figure \ref{ses:stru}. These laws of structural congruence have a strong correspondence to those for processes in Figure \ref{cong}. Because the equi-recursive equation $\mu \mathbf{t}. S\equiv S\{\mu \mathbf{t}. S/\mathbf{t}\}$ is left out, the session structural congruence is also decidable.

The operational semantics of sessions are presented in Figure \ref{ses:ops}, where we use $\lambda$ to denote $p,q:v$ or $\tilde{p}:B$. [\textsc{S-com}] describes the ordinary message-based communications between two participants. [\textsc{S-sess}] is for the session establishment and nesting; when a session is established, it runs interleavingly with its nested session. Other semantic rules are standard. [\textsc{S-times}], [\textsc{S-sum}], and [\textsc{S-con}] handle the session production, summation, and concatenation, respectively. Recursive sessions are dealt with by [\textsc{S-rec}] and session equivalence by [\textsc{S-eq}].

\begin{figure}[t]\vspace{-1em}
\begin{gather*}
   \hbox{[\textsc{S-com}]}~~\langle p,q:v\rangle \rightarrow B \overset{p,q:v }{\longrightarrow} B \qquad
   \hbox{[\textsc{S-sess}]}~~\langle \tilde{p}:B\rangle \{ A\}\overset{\tilde{p}:B}{\longrightarrow}A\otimes B\langle \tilde{p}\rangle \\
   \hbox{[\textsc{S-times}]}~~ \dfrac{S\overset{\lambda }{\longrightarrow} S'}{S\otimes T\overset{\lambda}{\longrightarrow}S'\otimes T} \qquad
   \hbox{[\textsc{S-sum}]}~~\dfrac{S\overset{\lambda }{\longrightarrow} S'}{S\oplus T\overset{\lambda}{\longrightarrow}S'} \qquad
   \hbox{[\textsc{S-con}]}~~\dfrac{S\overset{\lambda}{\longrightarrow} S'}{S ; T\overset{\lambda}{\longrightarrow}S';T}\\
   \hbox{[\textsc{S-rec}]}~~ \dfrac{S\overset{\lambda}{\longrightarrow} S'}{\mu \mathbf{t}.S \overset{\lambda}{\longrightarrow} S'\{\mu \mathbf{t}.S/\mathbf{t}\} }\qquad
  \hbox{[\textsc{S-eq}]}~~ \dfrac{S\equiv  T\quad T\overset{\lambda}{\longrightarrow}T' \quad T'\equiv S'}{S\overset{\lambda}{\longrightarrow}S'}
\end{gather*}\vspace{-0.6em}
\caption{Session operational semantics}\label{ses:ops}
\end{figure}

\subsection{Examples of Sessions}\label{ses:sessionexample}

We use the syntax of two-level sessions to formulate our business protocol. \hbem{Proto} is the session at the integrating level whilst the remaining four are at the communicating level. Their intuitive explanations have already been given in the last part of Sect.~\ref{sec:calculus}. We can also find a correspondence between \hbem{Proto} and its rough description in Introduction. $\bigoplus$ is the multiple case of $\oplus$.
\begin{align*}
    \hbem{Proto}  \overset{\mathrm{def}}{=} ~& \langle \mathrm{broker}, \mathrm{buyer}_1,\mathrm{buyer}_2:\hbem{Auction}\rangle \{\};\bigoplus_{i\in \{1,2\}} \langle \mathrm{buyer}_i,\\
    & \quad \mathrm{broker},\mathrm{seller}:\hbem{DTransaction}\rangle\{\}\;\oplus \langle \mathrm{buyer}_i,\mathrm{broker},\\
    & \qquad\mathrm{seller}:\hbem{STransaction}\rangle\{\langle \mathrm{buyer}_i,\mathrm{broker},\mathrm{bank}:\hbem{EPay}\rangle\{\}\} \\
    \hbem{Auction} \overset{\mathrm{def}}{=} ~& \bigoplus_{i\in \{2,3\}} \langle i,1: \mathrm{bid}\rangle \rightarrow \mu \mathbf{t}.\langle 1,5-i:\mathrm{quote}\rangle \rightarrow \\
    &\quad (\langle 1,i:\mathrm{invoice} \rangle \rightarrow\mathtt{end}
\oplus \langle 5-i,1:\mathrm{bid}\rangle \rightarrow \langle 1,i:\mathrm{quote}\rangle \\
    &\qquad\rightarrow (\langle i,1:\mathrm{bid}\rangle \rightarrow \mathbf{t} \oplus \langle 1,5-i:\mathrm{invoice}\rangle \rightarrow \mathtt{end})) \\
    \hbem{DTransaction} \overset{\mathrm{def}}{=}~& \langle 2,3:\mathrm{price}\rangle \rightarrow \langle 1,3:\mathrm{payment}\rangle \rightarrow \langle 3,1:\mathrm{order}\rangle \rightarrow \mathtt{end} \\
    \hbem{STransaction} \overset{\mathrm{def}}{=} ~& \langle 2,3:\mathrm{prepaid}\rangle \rightarrow \langle 3,1:\mathrm{order}\rangle \rightarrow\langle 1,2:\mathrm{confirm}\rangle \\
    & \quad \rightarrow \langle 2,3:\mathrm{payment}\rangle \rightarrow \mathtt{end}\\
    \hbem{EPay} \overset{\mathrm{def}}{=} ~& \langle 1,3:\mathrm{amount}\rangle \rightarrow \langle 3,2:\mathrm{transfer}\rangle \rightarrow \mathtt{end}
\end{align*}

A comparison of the above session formulation and the behavoiural formulation of the five agents in Sect.~\ref{sec:calculus} leads us to see three advantages of session modularisation and integration. First, sessions characterise the interactions between processes globally, facilitating the prevention of deadlocks and race conditions. Also, sessions are free of channels. Finally, following the principle of separation of concerns, communicating sessions partition protocols into independent modules whilst integrating sessions assemble communicating sessions at an adequately abstract level. The last aspect is unique to our theory and its merits are two-fold: it fits the natural understanding of the protocol and the gradual procedure of protocol formulation.

\subsection{Role Projection}\label{sec:roleprojection}

\emph{Session roles} or \emph{roles} refer to behaviours of participants acting in sessions. In other words, roles are the local description of sessions for participants. Formally, roles are represented as abstract processes. The goal of this sub-section is to develop mechanisms to project communicating and integrating sessions into their roles. The projection forms a basis for the type system developed later.

We first deal with the projection of communicating sessions.
First, we mark \emph{each} occurrence of \emph{each} event prefix in a given session by a unique channel name. Then, we map the given session into processes according to the following rules: 
\begin{itemize}
  \item $\mathtt{end}{\upharpoonright}r=\mathbf{0}$, $\mathbf{t}{\upharpoonright}r=X_\mathbf{t}$, $(\mu \mathbf{t}. B){\upharpoonright}r=\rec{X_\mathbf{t}} (B{\upharpoonright}r)$,
  \item $(\langle p,q:v\rangle\rightarrow B){\upharpoonright}r =$
  $$\left\{
      \begin{array}{ll}
        c!v. (B{\upharpoonright}r)  & \hbox{ if } r=p\\
        c?v. (B{\upharpoonright}r)  & \hbox{ if } r=q\\
        B{\upharpoonright}r  & \hbox{ if } r\neq p\neq q
      \end{array}
    \right.
   $$
  where the leftmost occurrence of $\langle p,q:v\rangle$ in $\langle p,q:v\rangle\rightarrow B$ is marked by $c$,
  \item $(B;B'){\upharpoonright}r=(B{\upharpoonright}r)\{B'{\upharpoonright}r/\mathbf{0}\}$, $(B\oplus B'){\upharpoonright}r=B{\upharpoonright}r+ B'{\upharpoonright}r$, and  $(B\otimes B'){\upharpoonright}r=B{\upharpoonright}r\2 B'{\upharpoonright}r$.
\end{itemize}

After the first two steps, we obtain a set of processes such that the \emph{message flow} between them at the runtime (according to the operational semantics) is \emph{deterministic} (and so channel interference is avoided). We say the message flow between the $P_1$ to $P_m$ deterministic if $P_1\2\ldots \2 P_m$ is deterministic. However, the number of channels used in sessions many be large. In the third step, we apply a channel substitution to the set of roles to optimise channel usage. The definition of the channel substitution is subject to practical considerations. For example, one may let two agents use the same channel to communicate, just as we did for the processes of five agents in the protocol example. The channel substitution is \emph{legal} as long as the message flow between the resulted processes remains deterministic. Without confusion, when writing $B{\upharpoonright}r$, we always refer to the optimised $B{\upharpoonright}r$ and call it the role of $B$ for $r$. We can check that the projection is well-defined based on the well-formedness of sessions.

The projection for integrating sessions is similar, but usually the number of communicating sessions in an integrating session is not very large, therefore we omit the channel optimisation step. First, we mark each \emph{occurrence} of the prefix in a given integrating session by a unique channel name. Then, we map the given session into processes according to the following rules: 
\begin{itemize}
  \item $\mathtt{end}{\upharpoonright}r=\mathbf{0}$, $\mathbf{t}{\upharpoonright}r=X_\mathbf{t}$, $(\mu \mathbf{t}. A){\upharpoonright}r =\rec{X_\mathbf{t}} (A{\upharpoonright}r)$,
  \item $\langle \tilde{p}:B\rangle \{A\}{\upharpoonright}r=$
  $$\left\{
      \begin{array}{ll}
        \bar{a}_{[2..n]}(\tilde{c}). (A{\upharpoonright}r) & \hbox{ if } r= \tilde{p}[1]\wedge \mathrm{pid}(B)=|\tilde{p}|=n ~\wedge \\
    & ~~\, |\mathrm{fc}(B{\upharpoonright}r)| =|\tilde{c}|\wedge \tilde{c}\cap \mathrm{fc}(A{\upharpoonright}r)=\varnothing \\
        a_{[k]}(\tilde{c}). (A{\upharpoonright}r)& \hbox{ if } r=\tilde{p}[k] \wedge |\mathrm{fc}(B{\upharpoonright}r)| =|\tilde{c}|\wedge \tilde{c}\cap \mathrm{fc}(A{\upharpoonright}r)=\varnothing  \\
        A{\upharpoonright}r & \hbox{ if } r\notin \tilde{p}
      \end{array}
    \right.
   $$
   where the leftmost occurrence of $\langle \tilde{p}:B\rangle$ in $\langle \tilde{p}:B\rangle \{A\}$ is marked by $a$,
  \item $(A;A'){\upharpoonright}r=(A{\upharpoonright}r)\{A'{\upharpoonright}r/\mathbf{0}\}$, $(A\oplus A'){\upharpoonright}r =A{\upharpoonright}r+ A'{\upharpoonright}r$,
  and $(A\otimes A'){\upharpoonright}r=A{\upharpoonright}r\2 A'{\upharpoonright}r$.
\end{itemize}The process $A{\upharpoonright}r$ is the role of $A$ for $r$.

The projection is completely automated for integrating sessions, but it presupposes a legal channel substitution for communicating sessions to optimise the channel usage.

\paragraph{\textbf{Examples of roles}} The following set of processes contains roles of $\hbem{Proto}$ for five agents (the first five) and roles of all four communicating sessions for the broker (the last four). Let $j\in \{1,2\}$.
\begin{align*}
    R_\mathrm{broker}^\mathrm{all} \overset{\mathrm{def}}{=} ~ & \overline{\mathrm{auc}}_{[2..3]}(a_{1,2},a_{1,3})\ldot \sum_{i\in \{1,2\}}(\mathrm{dTran}_{[2]}^i(b_{1,3},b_{2,3})\ldot\mathbf{0}\; + \\
    & \quad \mathrm{sTran}_{[2]}^i(c_{1,2},c_{1,3},c_{2,3})\ldot\mathrm{epay}_{[2]}^i(d_{1,3},d_{2,3})\ldot\mathbf{0}) \\
    R^\mathrm{all}_{\mathrm{buyer}_j} \overset{\mathrm{def}}{=} ~& \mathrm{auc}_{[j+1]}(a_{1,2},a_{1,3})\ldot (\overline{\mathrm{dTran}}_{[2..3]}^j(b_{1,3},b_{2,3})\ldot \mathbf{0}~+ \\
    & \quad  \overline{\mathrm{sTran}}_{[2..3]}^{j}(c_{1,2},c_{1,3}, c_{2,3})\ldot\overline{\mathrm{epay}}_{[2..3]}^{j}(d_{1,3},d_{2,3})\ldot \mathbf{0})\\
    R^\mathrm{all}_{\mathrm{seller}}\overset{\mathrm{def}}{=}~& \sum_{i\in \{1,2\}}( \mathrm{dTran}_{[3]}^i(b_{1,3},b_{2,3})\ldot \mathbf{0} +\mathrm{sTran}_{[3]}^j(c_{1,2},c_{1,3},c_{2,3})\ldot\mathbf{0}) \\
    & \quad  \overline{\mathrm{sTran}}_{[2..3]}^{i-1}(c_{1,2},c_{1,3}, c_{2,3})\ldot\overline{\mathrm{epay}}_{[2..3]}^{i-1}(d_{1,3},d_{2,3})\ldot \mathbf{0}) \\
    R^{\mathrm{all}}_{\mathrm{bank}} \overset{\mathrm{def}}{=} ~& \sum_{i\in\{1,2\}} \mathrm{epay}_{[3]}^i(d_{1,3},d_{2,3})\ldot \mathbf{0}\\
    R_\mathrm{broker}^\mathrm{auc} \overset{\mathrm{def}}{=} ~& \sum_{i\in \{1,2\}} (a_{1,i+1}?\mathrm{bid}\ldot\rec{X_1}(a_{1,4-i}!\mathrm{quote}\ldot(a_{1,i+1}!\mathrm{invoice}\ldot\mathbf{0}\ldot +\\
    &  \quad a_{1,4-i}?\mathrm{bid}\ldot a_{1,i+1}!\mathrm{quote}\ldot(a_{1,i+1}?\mathrm{bid}\ldot X_1+a_{1,4-i}!\mathrm{invoice}\ldot\mathbf{0})))) \\
    R_\mathrm{broker}^\mathrm{sTran} \overset{\mathrm{def}}{=} ~&  c_{2,3}!\mathrm{prepaid}\ldot c_{1,2}?\mathrm{confirm}\ldot c_{2,3}!\mathrm{payment}\ldot\mathbf{0}\\
    R_\mathrm{broker}^\mathrm{dTran} \overset{\mathrm{def}}{=} ~& b_{2,3}!\mathrm{price}\ldot\mathbf{0} \qquad\qquad R_\mathrm{broker}^\mathrm{epay} \overset{\mathrm{def}}{=} d_{2,3}?\mathrm{transfer}\ldot\mathbf{0}
\end{align*}

\section{Type Discipline for Sessions} \label{sec:typediscipline}

\subsection{Type System}\label{sec:typesystem}

\begin{figure}[t]\vspace{-1em}
\begin{gather}
   ~~~~~~~~~~~~~~~~~~\Gamma \vdash \mathbf{0}\triangleright \mathbf{0} \qquad \Gamma, a\triangleright  B \vdash a\triangleright  B \tag*{\textsc{[T-nil]},\textsc{[T-ch]}}\label{ty:t-nil-ch}\\
   \Gamma, X\vdash X\triangleright X \hbox{ or }  \Gamma, X\vdash \mathbf{0}\circ\tilde{c}:X \tag*{\,[\textsc{T-var}]}\label{ty:t-var}\\
   \dfrac{\Gamma\vdash a\triangleright  B \quad \Gamma\vdash P\triangleright R\circ \tilde{c}:B{\upharpoonright} 1\langle \tilde{c}\rangle, \Delta\quad \mathrm{pid}(B)=[1,n] }
   {\Gamma \vdash \bar{a}_{[2..n]}(\tilde{c}).P\triangleright \bar{a}_{[2..n]}(\tilde{c}).R\circ \Delta} \tag*{\textsc{[T-inv]}}\label{ty:t-inv}\\
   \dfrac{\Gamma\vdash a\triangleright B \quad \Gamma\vdash P\triangleright R\circ\tilde{c}:B{\upharpoonright}i\langle \tilde{c}\rangle, \Delta\quad  2\leq i\in \mathrm{pid}(B)}
   {\Gamma \vdash a_{[i]}(\tilde{c}).P\triangleright a_{[i]}(\tilde{c}).R\circ \Delta} \tag*{\textsc{[T-acc]}}\label{ty:t-acc}\\
   \dfrac{\Gamma \vdash P\triangleright R\circ \Delta \quad \tilde{c}\cap (\mathrm{dom}(\Gamma)\cup  \mathrm{ch}(\Delta))=\varnothing}
   {\Gamma \vdash P \triangleright R\circ \tilde{c}:\mathbf{0},\Delta } \tag*{\textsc{[T-tml]}}\label{ty:t-tml} \\
   \dfrac{\Gamma \vdash P\triangleright R\circ \Delta \quad \tilde{c}\cap (\mathrm{dom}(\Gamma)\cup \mathrm{ch}(\Delta))=\varnothing}
   {\Gamma \vdash P \triangleright R\circ \Delta, \tilde{c}:\mathbf{0} } \tag*{\textsc{[T-tmr]}}\label{ty:t-tmr} \\
   \dfrac{\Gamma \vdash P\triangleright R\circ   \tilde{c}  :Q,\Delta \quad b\in \tilde{c}}
   {\Gamma \vdash b\S v.P\triangleright R\circ  \tilde{c}: b\S v.Q,\Delta}\quad \hbox{where } \S\in\{!,?\} \tag*{\textsc{[T-sr]}}\label{ty:t-sr} \\
   \dfrac{\Gamma \vdash P\triangleright R\circ \Delta \quad \Gamma \vdash P'\triangleright R'\circ \Delta' \quad\Delta\asymp \Delta'}{\Gamma\vdash P\2 P' \triangleright R\2 R'\circ \Delta\2 \Delta'}\tag*{\textsc{[T-com]}}\label{ty:t-com} \\
   \dfrac{\Gamma\vdash P\triangleright R\circ \Delta \quad \Gamma\vdash P'\triangleright R'\circ \Delta' \quad \Delta \asymp\Delta'}{\Gamma\vdash P+ P' \triangleright R\tsum R'\circ \Delta\tsum \Delta'}\tag*{\textsc{[T-sum]}}\label{ty:t-sum}\\
   \dfrac{\Gamma,  X \vdash P\triangleright R\circ \tilde{c}_1:Q_1,\ldots,\tilde{c}_n:Q_n}
   {\Gamma\vdash \rec{X} P\triangleright R^{[X]}\circ \tilde{c}_1: Q^{[X]}_1, \ldots, \tilde{c}_n: Q^{[X]}_n }\tag*{\textsc{[T-rec]}}\label{ty:t-rec}\\
   \dfrac{\Gamma\vdash P\triangleright R\circ \Delta \quad R\equiv R' \quad \Delta\equiv \Delta'}{\Gamma \vdash P\triangleright R'\circ \Delta'} \tag*{\textsc{[T-eq]}}\label{ty:t-eq}\\
\dfrac{\Gamma \vdash P\triangleright R\circ\tilde{c}:Q,\Delta \quad  b\in \tilde{c}}{\Gamma\vdash (\nu b) P\triangleright R\circ\Delta, \tilde{c}\backslash b:Q,\Delta'}\tag*{\textsc{[T-hid]}}\label{ty:t-hid}\\
\dfrac{\Gamma \vdash P\triangleright R\circ \Delta \quad b\not\in \mathrm{dom}(\Gamma)\cup \mathrm{ch}(\Delta)}{\Gamma \vdash (\nu b)P\triangleright R\circ \Delta}\tag*{\textsc{[T-vei]}}\label{ty:t-vei}\\
\dfrac{\Gamma \vdash P\triangleright R\circ \Delta}{\Gamma \vdash l: P\triangleright R\circ \Delta}\tag*{\textsc{[T-lab]}}\label{ty:t-lab}
\end{gather}\vspace{-0.8em}
\caption{Typing rules}\label{ty:rule}
\end{figure}

The purpose of the type system below is to efficiently type processes so that the `illegal' runtime behaviours of processes are prevented by static type checking. The type system is based on the role projection developed earlier.

We define the following syntax:
\begin{align*}
    \Gamma::=  \varnothing ~|~ \Gamma, a\triangleright S ~|~ \Gamma, X\qquad
    \Delta::=  \{\tilde{c}_i:Q_i\}_{i\in I}
\end{align*}

A \emph{type environment} $\Gamma$ is a function that assigns sessions to some channels (session channels) and \emph{typing} to session variables. A typing is of the form $R\circ \Delta$, where $R$, called a \emph{session typing}, is projected from an integrating session, and $\Delta$, called a \emph{channel typing}, is a sequence of processes labelled by disjoint channel sequences.
The domain of $\Gamma$ is a set of channels or variables it acts on. If its domain contains channel names only, we say $\Gamma$ is \emph{pure}.
Re-ordering of items in a type environment $\Gamma$ is permitted, but forbidden in a channel typing $\Delta$.

The
\emph{type judgement} $\Gamma\vdash P\triangleright R\circ \Delta$ reads `the typing of $P$ is $R\circ\Delta$ under $\Gamma$.' If $\Delta=\epsilon$, we write $\Gamma\vdash P\triangleright R$. Formally, type judgements are defined by the typing rules in Figures \ref{ty:rule}, which are explained later. We also say $P$ is \emph{typed} or \emph{typable} by $\Gamma$ if there is a typing of $P$ under $\Gamma$.
A few auxiliary definitions are given. $|\Delta|$ is the length of $\Delta$ and $\Delta[i]$ is the $i$th item of $\Delta$ where $1\leq i\leq |\Delta|$. $\Delta$ and $\Delta'$ are \emph{compatible}, denoted $\Delta \asymp \Delta'$, if $|\Delta|=|\Delta'|$ and $\Delta[i]$ and $\Delta'[i]$ have the same labelling sequence of channels for each $1\leq i\leq |\Delta|$.
Let $\Delta=\tilde{c}_1:Q_1,\ldots, \tilde{c}_n:Q_n$ and $\Delta'=\tilde{c}_1:Q_1',\ldots, \tilde{c}_n:Q_n'$. $\Delta\equiv\Delta'$ if $Q_i\equiv Q_i'$ for each $1\leq i\leq n$. Let $\Delta\2 \Delta'=\tilde{c}_1:Q_1\2 Q_1',\ldots,\tilde{c}_n:Q_n\2 Q_n'$ and $\Delta\tsum \Delta'=\tilde{c}_1:Q_1\tsum Q_1',\ldots, \tilde{c}_n:Q_n\tsum Q_n'$. $\mathrm{ch}(\Delta)$ is the set of all labelling channels in $\Delta$.

\ref{ty:t-inv} and \ref{ty:t-acc} are for session invitation and acceptance, and \ref{ty:t-sr} for the ordinary communication. \ref{ty:t-tml} and \ref{ty:t-tmr} are needed because $\Delta\asymp \Delta'$ is used in the pre-conditions of \ref{ty:t-com} and \ref{ty:t-sum}.
By \ref{ty:t-var}, the type variables happen in either the main (i.e.~integrating) session or a single communicating session. \ref{ty:t-rec} handles the recursive construction where $P^{[X]}$ is defined in Sect.~\ref{sec:calculus}. \ref{ty:t-eq} is necessary to make the current type system expressive enough but also bring in the infinity of typing for processes.  For restricted processes e.g.~$(\nu a) P$, if $a\in \mathrm{fv}(P)$, then it is dealt with by \ref{ty:t-hid}; otherwise, by \ref{ty:t-vei}. \ref{ty:t-lab} absorbs the labelling in the typing derivation. \ref{ty:t-nil-ch} are standard.

We construct a (pure) type environment for the five agents in the business protocol and establish type judgements for them.
\begin{pr1}\label{pr:typeexample}
Let
$
    \Gamma_{\mathrm{prt}}=\mathrm{auc}~\triangleright \hbox{Auction},~  \mathrm{dTran}~\triangleright\hbox{DTran},~
 \mathrm{sTran}~\triangleright \hbox{STran},$ $ \mathrm{epay}~\triangleright \hbox{EPay}
$.
We have that
$\Gamma_{\mathrm{prt}} \vdash P_\mathrm{broker} \triangleright R_\mathrm{broker}^\mathrm{all}$, $\Gamma_{\mathrm{prt}} \vdash P_{\mathrm{buyer}_i}\triangleright R_{\mathrm{buyer}_i}^\mathrm{all}$, $\Gamma_{\mathrm{prt}} \vdash P_\mathrm{seller} \triangleright R_\mathrm{seller}^\mathrm{all}$, and $\Gamma_{\mathrm{prt}} \vdash P_\mathrm{bank}\triangleright R_\mathrm{bank}^\mathrm{all}$.
\end{pr1}
\subsection{Properties of Typing}\label{sec:ty-sound}

We study several key properties of the type system. The decidability of type inference comes first.
\begin{th1}\label{th:typeinf}
Given a process $P$ and a type environment $\Gamma$, it is decidable whether there exist $R,\Delta$ such that $\Gamma \vdash P\triangleright R\circ \Delta$. If there exist, then there is an algorithm to construct such a pair.
\end{th1}

The proof of Theorem \ref{th:typeinf} is by computing a so-called \emph{principal} typing for a given process under some type environment. A principal typing is a particular typing for a process such that the process has the principal typing if and only if it is typable. A standard type checking algorithm can be constructed to (attempt to) compute the principal typing for each process, and the termination of the algorithm is guaranteed by the decidability of the structural congruence for processes (cf.~Lemma \ref{le:process-equivalence}).



To present the following three properties of the type system, we put forward an auxiliary definition: for a channel typing $\Delta=\tilde{c}_1:Q_1,\ldots, \tilde{c}_n:Q_n$, let $\lceil \Delta\rceil$ be the multiple parallel-composition process $ Q_1\2\ldots\2 Q_n$. In general, $\Delta \equiv \Delta'$ is strictly stronger than $\lceil \Delta\rceil\equiv \lceil \Delta'\rceil$.

The Subject Congruence Theorem below implies that if $P$ is typable and $P\equiv Q$ then $Q$ is also typable and their typing have a certain structural relation.
\begin{th1}[Subject congruence] \label{th:subcon}If $\Gamma\vdash P\triangleright R\circ \Delta$ and $P\equiv P'$, then there exist $R', \Delta'$ such that $\Gamma\vdash P'\triangleright R'\circ \Delta'$, $R \equiv  R'$ and $ \lceil \Delta \rceil \equiv \lceil \Delta'\rceil$.
\end{th1}

The Subjection Reduction Theorem states that the typability of a process is preserved in an `expected' way during its evolvement.
The theorem rules out the standard type errors. For example, there is no $\Gamma$ such that $\Gamma \vdash \bar{a}_{[2..n]}(\tilde{c}).P_1\2 a_{[n+1]}(\tilde{c}).P_2\2 P_3$ or $\Gamma \vdash a_{[k]}(\tilde{c}).Q_1\2 a_{[l]}(\tilde{c}').Q_2$ where $|\tilde{c}|\neq |\tilde{c}'|$.
\begin{th1}[Subject reduction]\label{th:subred} If $\Gamma\vdash P\triangleright R\circ \Delta$ and $P\overset{\alpha}{\longrightarrow}P'$, then $\Gamma\vdash P'\triangleright R'\circ \Delta'$ for some $R',\Delta'$ satisfying the following conditions:
\begin{enumerate}
  \item If $\alpha=a\S v$ where $\S\in \{!,?\}$ then $R \succ R'$ and $\lceil \Delta\rceil\overset{a \S v}{\longrightarrow}_\succ \lceil \Delta'\rceil$,
  \item If $\alpha=\bar{a}_{[2..n]}(\tilde{c})$ and $\Gamma\vdash a \triangleright B$ then $R \overset{\bar{a}_{[2..n]}(\tilde{c})}{\longrightarrow}_\succ  R'$ and $B{\upharpoonright}1\langle \tilde{c}\rangle \2 \lceil \Delta\rceil \succ \lceil \Delta' \rceil $,
  \item If $\alpha=a_{[k]}(\tilde{c})$ and $\Gamma\vdash a \triangleright B$ then $ R\overset{a_{[k]}(\tilde{c})}{\longrightarrow}_\succ  R'$ and $ B{\upharpoonright}k\langle \tilde{c}\rangle\2 \lceil \Delta\rceil\succ \lceil \Delta' \rceil $,
  \item If $\alpha=\tau$ then
\begin{enumerate}
    \item either $ R\succ R'$ and $\lceil \Delta\rceil\overset{\tau}{\longrightarrow}_\succ \lceil \Delta'\rceil$,
    \item or $ R\overset{\tau}{\longrightarrow}_\succ  R'$ and $B{\upharpoonright}1\langle \tilde{c}\rangle\2\ldots \2 B{\upharpoonright}n \langle \tilde{c}\rangle\2  \lceil \Delta\rceil\succ \lceil \Delta' \rceil $ for some $B, n$ such that $\mathrm{pid}(B)=[1,n]$.
  \end{enumerate}
\end{enumerate}
\end{th1}

The last property says that the typing of a process under a type environment is unique up to a certain structural relation as in Theorem \ref{th:subcon}.
\begin{th1}[Typing uniqueness]\label{th:typcon}
If $\Gamma\vdash P\triangleright R\circ \Delta$ and $\Gamma\vdash P\triangleright R'\circ \Delta'$, then $R\equiv R'$ and $\lceil \Delta\rceil\equiv \lceil \Delta'\rceil$.
\end{th1}

\section{Behavioural Analysis}\label{sec:analysis}

In this section, we use the two-level session types to analyse interactions of distributed program components. Two system properties are dealt with: a channel safety property (also studied in the existing session type literature) and a behavioural conformance between processes and sessions. But we first show how to represent and properly type a distributed system in our formalism.

Informally, a program or a component in a distributed system is a pair of a participant name and a process. Following works in the process algebraic approach to architectural analysis, such as \cite{ag97,bcd02,syz12}, we define (the architecture of) a system as a parallel composition of programs or components. Formally, we define that
\begin{df1}[Systems]\label{df:intf}\label{df:arch}
A program or a component is a labelled process $r:P$, where $r$ and $P$ specify its name and behaviour, respectively.
A system is a process of the form
$\mathrm{Sys} = r_1:P_1 \2 \ldots\2 r_n:P_n$.
\end{df1}

For example, the following system implements the business protocol introduced in the Introduction and formalised in Sect.~\ref{ses:sessionexample}.
\begin{align*}
    \mathrm{Sys}_e ~=~ \,& \mathrm{broker}:P_{\mathrm{broker}}\2 \mathrm{buyer}_1:P_{\mathrm{buyer}_1}\2 \\
    & \quad \mathrm{buyer}_2:P_{\mathrm{buyer}_2} \2 \mathrm{seller}:P_\mathrm{seller}\2 \mathrm{bank}:P_\mathrm{bank}
\end{align*}

A session channel $a$ \emph{marks} $B$ in $A$ if \emph{some} occurrence of $B$ in $A$ is marked by $a$ in the projection. The following definition characterises how to use session types to properly type a system.

\begin{df1}[Session well-typedness]
$\mathrm{Sys}$ (as defined in Def.~\ref{df:arch}) is well-typed by $A_\mathrm{spc}$ under $\Gamma$ if
\begin{itemize}
  \item $\mathrm{pid}(A_\mathrm{spc})=\{r_1,\ldots,r_n\}$,
  \item $\Gamma \vdash a\triangleright B$ if and only if $a$ marks $B$ in $A_\mathrm{spc}$, and
  \item $\Gamma \vdash P_i\triangleright A_\mathrm{spc}{\upharpoonright}r_i $ for each $1\leq i\leq n$.
\end{itemize}
\end{df1}

If $\mathrm{Sys}$ is well-typed by $A_\mathrm{spc}$, we call $A_\mathrm{spc}$ a session for $\mathrm{Sys}$.
In general, well-typedness is strictly stronger than typability. In other words, if $\mathrm{Sys}$ is well-typed by $A_\mathrm{spc}$ under $\Gamma$ then $\Gamma \vdash \mathrm{Sys}\triangleright A_\mathrm{spc}$; but the other direction does not necessarily hold. Also, we observe that if $\mathrm{Sys}$ is well-typed by some session and $\mathrm{Sys}$ (as a process) is a close then $\mathrm{Sys}$ is well-typed by some pure session.

The channel safety property below says that channel interference is prevented at the runtime of the system.
\begin{df1}[Channel privacy] The communicating channels in $\mathrm{Sys}$ are private if the following holds:
if $\mathrm{Sys}\overset{\tau}{\longrightarrow}_\ast (\nu \tilde{b})(r_1:P_1\2 \ldots \2 r_n:P_n)$ and
$c\S v \in $ $\mathrm{act}(P_i)$ where $\S \in\{!,?\}$, then their exists a unique $r_j$ such that $r_i\neq r_j$ and $c\in\mathrm{fc}(P_j)$.
\end{df1}
The channel privacy of a system is a consequence of well-typedness by a session specification, as the following theorem demonstrates.
\begin{th1}\label{th:chanpri}
If $\mathrm{Sys}$  is well-typed by $A$ under $\Gamma$, then the communicating channels in $\mathrm{Sys}$ are private.
\end{th1}
Informally, the theorem is guaranteed by the determinism of message flow in the projected roles and the creation of fresh channels in the session establishment.

Session conformance says that the runtime interactions of the system conform to its session specification.
\begin{df1}[Session conformance]\label{df:conformance}  $P$ conforms to $S$, if there is a relation $\mathcal{R}$ of processes and sessions such that if $\langle P,S\rangle\in \mathcal{R}$ then the following conditions hold:
\begin{itemize}
  \item If $P \equiv (\nu \tilde{b}) (p:Q_1\2 q:Q_2\2 R )$, $Q_1\overset{a!v}{\longrightarrow} Q_1'$ and $Q_2\overset{a?v}{\longrightarrow} Q_2'$, then there exists $S'$ such that $S\overset{p,q:v}{\longrightarrow} S'$ and $\langle P', S'\rangle \in \mathcal{R}$ where $P'\equiv (\nu \tilde{b}) (p:Q_1'\2 q:Q_2'\2 R)$;
  \item If $P\equiv (\nu \tilde{b}) (p_1: Q_1\2 \ldots \2 p_m:Q_{m}\2 R)$, $Q_1\overset{\bar{a}_{[2..m]}(\tilde{c})}{\longrightarrow}Q_1'$ and $Q_i\overset{a_{[i]}(\tilde{c})}{\longrightarrow}Q_i'$ for all $2\leq i\leq m$, then there exist $B,S'$ such that $\Gamma\vdash a\triangleright B$,  $S\overset{p_{1},\ldots,p_m: B}{\longrightarrow} S'$ and $\langle P',S'\rangle\in \mathcal{R}$, where $P'\equiv (\nu \tilde{b}) ((\nu \tilde{c}) (p_1:Q_1'\2\ldots \2 p_m:Q_m')\2 R) $.
\end{itemize}
\end{df1}
An alternative explanation of session conformance is behavoural refinement, because Def.~\ref{df:conformance} actually defines a behavioural stimulation relation between sessions and processes. We observe that if $P$ conforms to $S$ and $P\equiv P'$, then $P'$ conforms to $ S$. The following theorem confirms that session conformance of the system is also a consequence of well-typedness by a session specification.
\begin{th1}\label{th:sesfid} If $\mathrm{Sys}$ is well-typed by $A_\mathrm{spc}$ then $\mathrm{Sys}$ conforms to $ A_\mathrm{spc}$.
\end{th1}

By what we have established so far, we have the following two properties for the system $\mathrm{Sys}_e$:
\begin{pr1}
(1) The communicating channels are private in $\mathrm{Sys}_e$. (2) The behaviour of $\mathrm{Sys}_e$ conforms to its session specification $\hbox{Proto}$.
\end{pr1}

\section{Process Slicing}\label{sec:slicing}

A type inference algorithm computes a typing, if possible, for a process under a type environment (cf.~Theorem \ref{th:typeinf}). However, in the real-life cases, the developers have the session specification in the first place and then implement it, so they need to check whether a process is typable by the given session specification. A straightforward method to solve this type checking problem consists of two steps: to verify whether $\Gamma\vdash P\triangleright A{\upharpoonright}r$ for some $r:P,A, \Gamma$, we first compute $\Gamma\vdash P\triangleright A_P{\upharpoonright}r$ by a type inference algorithm and then check whether $A_P\equiv A$. This algorithm is efficient, pre-supposing we have an efficient type inference algorithm.

However, there is a drawback in the above algorithm: if $\Gamma\vdash P\triangleright A{\upharpoonright}r$ does not hold, the algorithm does not tell which session or sessions it violates. Since the session specification is modularised, it is desirable to know the violated session or sessions. In the following, we propose an algorithm based on process slicing to improve the type checking. Informally, the key idea of the algorithm is to decompose a process into parts and compare each part with a role projected from a corresponding session.

Suppose each session channel in $P$ is typed by $\Gamma$, namely, contained in the domain of $\Gamma$. The algorithm consists of two steps.
The first step is the process slicing. Because the hiding and labelling operators are unnecessary for processes as the initial (not runtime) behaviours of programs or components, we assume that $P$ is free of these two operators. We call $P^{\chi_M}$ the main slice of $P$ and the main slicing function $\chi_M$ is formally defined as follows.
\begin{align*}
    \mathbf{0}^{\chi_M} = & ~\mathbf{0} \quad X^{\chi_M}=X \quad (\rec{X} P)^{\chi_M} = \rec{X} (P)^{\chi_M}, \\
    (\pi.P)^{\chi_M}  = &~
\left\{
  \begin{array}{ll}
     \pi.(P)^{\chi_M} & \hbox{ if $\pi$ is a session action,} \\
     P^{\chi_M} & \hbox{ otherwise;}
  \end{array}
\right.  \\
    (P \star Q)^{\chi_M} = & ~ P^{\chi_M}\star Q^{\chi_M} \quad  \hbox{where $\star$ is $\2$ or $+$.}
\end{align*}

We call $P^{\chi_{\tilde{c}}}$ the $\tilde{c}$-slice of $P$, and the slicing function $\chi_{\chi_{\tilde{c}}}$, which are parametric on $\tilde{c}$, is defined below.
\begin{align*}
    \mathbf{0}^{\chi_{\tilde{c}}} = & ~\mathbf{0} \quad X^{\chi_{\tilde{c}}}=X \quad (\rec{X} P)^{\chi_{\tilde{c}}} = \rec{X} (P)^{\chi_{\tilde{c}}}, \\
    (\pi.P)^{\chi_{\tilde{c}}} = & ~
\left\{
  \begin{array}{ll}
    P^{\chi_{\tilde{c}}} & \hbox{ if } \mathrm{fc}(\pi)\not\in \tilde{c}, \\
    \pi.(P)^{\chi_{\tilde{c}}} & \hbox{ otherwise;}
  \end{array}
\right.
 \\
    (P \star Q)^{\chi_{\tilde{c}}} = & ~ P^{\chi_{\tilde{c}}}\star Q^{\chi_{\tilde{c}}} \quad  \hbox{where $\star$ is $\2$ or $+$.}
\end{align*}

After computing the slices of a process, we check whether each slice is structurally congruent to a corresponding role. Specifically, we verify if $A{\upharpoonright}r\equiv (\pi.P)^{\chi_M}$, $B{\upharpoonright}1\langle \tilde{c}\rangle \equiv P^{\chi_{\tilde{c}}}$, and $B{\upharpoonright}k\langle \tilde{c}\rangle \equiv P^{\chi_{\tilde{c}}}$, where $\Gamma\vdash a\triangleright B$.

By by our bound name convention, a name is not bound twice and does not have free \emph{and} bound occurrences simultaneously in a process.
The following theorem says that if $P$ is typed by $A{\upharpoonright}r$ under $\Gamma$ then the slicing of $P$ `coincides' with the role projection of $A$.
\begin{th1}[Slicing-projection correspondence] \label{th:slicing} If
$\Gamma \vdash P\triangleright A{\upharpoonright}r\langle \tilde{a}\rangle$ then the following three conditions hold:
\begin{itemize}
  \item $A{\upharpoonright}r\langle \tilde{a}\rangle \equiv P^{\chi_M}$,
  \item if $\bar{a}_{[2..n]}(\tilde{c})\in \mathrm{act}(P)$ and $\Gamma\vdash a\triangleright B$, then $B{\upharpoonright}1\langle \tilde{c}\rangle \equiv  P^{\chi_{\tilde{c}}}$,
  \item if $a_{[k]}(\tilde{c})\in \mathrm{act}(P)$ and $\Gamma\vdash a\triangleright B$, then $B{\upharpoonright}k\langle \tilde{c}\rangle \equiv  P^{\chi_{\tilde{c}}}$.
\end{itemize}
\end{th1}

Based on the above theorem, the correctness of the process slicing algorithm for the type checking is established. However, the method is not complete: the other direction of the theorem does not hold, as witnessed by the following counter-example. Thereby, the coincidence of role projection and process slicing does not entail the typability, and a technical implication is that the process slicing method cannot replace the type system in Sect.~\ref{sec:typesystem}.

\begin{pr1}Let $B_1=\langle p,q:v_1\rangle\rightarrow \langle p,q:u_1\rangle \rightarrow\mathtt{end}$, $B_2=\langle q,p:v_2\rangle \rightarrow \langle q,p:u_2\rangle \rightarrow \mathtt{end}$, $A_0=B_1;B_2$, $\Gamma_0 \vdash a_i\triangleright B_i$ where $i\in \{1,2\}$, and $P_1=a^1_{[2]}(c_1).\, $ $c_1?v_1.\,  a^2_{[2..2]}(c_2).\,  c_2!v_2.\, c_1?u_1.\,  c_2!u_2.\,  \mathtt{end}$. With a suitable role projection of $B_1$ and $ B_2$, we have that $A_0, P_1$ and $\Gamma_0$ satisfy the three conditions in Theorem \ref{th:slicing} but not $\Gamma_0\vdash P_1\triangleright A_0{\upharpoonright}p\langle a_1,a_2\rangle$.
\end{pr1}

\section{Related Work}\label{sec:relatedwork}

\paragraph{\textbf{Session type theories}}

Our work is rooted in the forgoing theories of session types, especially the global description of interactions and multiparty sessions. Carbone \etal \cite{chy07} presented two calculi to describe the communication behaviours from the global and local perspectives, respectively, and several principles to establish a sound and complete projection of the former to the latter. Some of the ideas behind the syntactic restrictions that we set up for the two-level sessions are related to their projection principles. The process calculus in the present paper is from Honda \etal \cite{hyc08}, in which the authors extended the traditional binary session types to the multiparty asynchronous context and solved several technical channels (as the result of the loss of two-party duality and the asynchrony) such that several fundamental properties of the session type discipline also hold by linearity analysis. The syntax for the calculus is abstract (e.g.~messages are treated as message types) and does not contain some syntactic features that are considered as essential to session type theories (e.g.~the distinction of internal and external choices and message-based branching behaviours, as argued by Castagna1 and Padovani \cite{cp09}). Our intention is to focus on the two-level separation of session syntax and minimise the side techniques when studying relevant properties. We leave the  work on enriching the syntax of session and calculi alike according to the existing session type theories in the future.


The subsequent work on session types witnesses a trend of increment on the expressive power to characterise richer conversation structures. For example,  Deni\'{e}lou and Yoshida \cite{dy11} extended the multiparty session types to accommodate the runtime change of session participants, i.e.~the joining or leaving of participants, after a session is initiated. The same authors \cite{ydbh10} introduced a finite recursive type constructors into the multiparty session types to express a wide range of processes whose specification structures are parameterised whilst keeping the type checking for the resulting type system decidable. To improve protocol modularisation of session types, Demangeon and Honda \cite{dh12} introduced a way to define abstract nested protocols independent of their host protocols such that the host protocols can call the nested ones by passing them arguments such as values, roles, and even (names of) other protocols. In these studies, the enrichment of the session type construction leads to the increment of syntactic primitives in the process calculi. In contrast, the separation of two-level sessions in our work does not complicate the syntax of the calculus. An interesting point is to compare the concept of nested protocols by Demangeon and Honda \cite{dh12} with that in the present paper. Their protocol calling is comparable to the procedure calling in the sequential programming, in which the exact position of the involvement must be specified to make sense of the main program. Our protocol nesting is more general in the sense that `being nested by' just means `occurring within'. Also, in our work, the meaning of the host protocol is complete with or without its nested protocol(s).

Padovani \cite{pad12} proposed a backward approach to session types, in which session types are defined as projected fragments of processes. More specifically, a process is sliced as per channels it uses and session types are a type approximation of the channel-sliced fragments of the process. There are two connecting points between his work and ours: first, both make use of process slicing, in spite of different purposes; second, both (and \cite{chy07}) investigate sessions semantically.

\paragraph{\textbf{Session types as architectural connection}}

The idea of viewing sessions as a behavioural approximation of processes comes from process algebraic analysis of software architectural connection. Architectural connection deals with the interactions of components which contrast to the local computations of components. Allan and Garlan \cite{ag97} argued for the merits of implementing architectural connection in a special class of components called connectors. They formulated connector types based on the process algebra \mbox{CSP} \cite{hoa78} and analysed the protocol compatibility issues related to components and connectors. Following their approach, the present authors \cite{syz10}\cite{syz12} proposed formal languages and methods to improve the architectural analysis. But these works assume the co-ordination of connectors for components and, hence, only handle the connector-based architectural styles. Bernardo \etal \cite{bcd02} distinguished the connector-based and non-connector-based styles, but their analytic techniques for the latter are based on the notion of `inter-operability' of a process against others, which skirts around the problem. Multiparty session types offer a solution to overcome the restriction by describing the component \mbox{interactions} globally without using connectors. To employ multiparty session types to analyse architectural connection, we need to be concerned with the behavioural compatibility (defined as session conformance in the present paper) between component computations (processes) and their expected interactions (session types).


\section{Conclusions}

We address the problem of session integration in protocol specification and develop a theory of two-level synchronous multiparty session types, in which session integration is separated from session communication. As of the technical results, we develop a new type system and study its key properties. We also analyse a channel safety property and a behavioural relation between processes and sessions, and present a process slicing method to improve the type checking.

We outline several interesting directions for further studies. First, we are working on the analysis of more behavioural properties of distributed computing systems in the novel session type theory. For example, behavioural refinement is two weak in some sense, and we want to establish a relation between behavioural refinement and equivalence between processes and sessions. Deadlock-freedom and liveness of processes are also important properties to be studied. The challenge is to properly revise the set of typing rules so that the satisfaction of some natural properties by the sessions entails the satisfaction of those behavioural properties. Second, we also expect to enrich the syntax of our process calculus according to existing session type studies. Third, the process slicing method is not complete with respect to the type system and, therefore, one research question revolves around finding a complete method to facilitate the type checking. Finally, we are also interested in leveraging session types as a theoretic tool for software architectural analysis.


\newpage

\section*{Appendix}

\appendix

\section{Complete Set of Roles for The Protocol Example}

\paragraph{\textbf{Roles of Proto for the five agents}} Let $j\in \{1,2\}$.
\begin{align*}
    R_\mathrm{broker}^\mathrm{all} \overset{\mathrm{def}}{=} ~ & \overline{\mathrm{auc}}_{[2..3]}(a_{1,2},a_{1,3})\ldot \sum_{i\in \{1,2\}}(\mathrm{dTran}_{[2]}^i(b_{1,3},b_{2,3})\ldot\mathbf{0}\; + \\
    & \quad \mathrm{sTran}_{[2]}^i(c_{1,2},c_{1,3},c_{2,3})\ldot\mathrm{epay}_{[2]}^i(d_{1,3},d_{2,3})\ldot\mathbf{0}) \\
    R^\mathrm{all}_{\mathrm{buyer}_j} \overset{\mathrm{def}}{=} ~& \mathrm{auc}_{[j+1]}(a_{1,2},a_{1,3})\ldot (\overline{\mathrm{dTran}}_{[2..3]}^j(b_{1,3},b_{2,3})\ldot \mathbf{0}~+ \\
    & \quad  \overline{\mathrm{sTran}}_{[2..3]}^{j}(c_{1,2},c_{1,3}, c_{2,3})\ldot\overline{\mathrm{epay}}_{[2..3]}^{j}(d_{1,3},d_{2,3})\ldot \mathbf{0})\\
    R^\mathrm{all}_{\mathrm{seller}}\overset{\mathrm{def}}{=}~& \sum_{i\in \{1,2\}}( \mathrm{dTran}_{[3]}^i(b_{1,3},b_{2,3})\ldot \mathbf{0} +\mathrm{sTran}_{[3]}^j(c_{1,2},c_{1,3},c_{2,3})\ldot\mathbf{0}) \\
    & \quad  \overline{\mathrm{sTran}}_{[2..3]}^{i-1}(c_{1,2},c_{1,3}, c_{2,3})\ldot\overline{\mathrm{epay}}_{[2..3]}^{i-1}(d_{1,3},d_{2,3})\ldot \mathbf{0}) \\
    R^{\mathrm{all}}_{\mathrm{bank}} \overset{\mathrm{def}}{=} ~& \sum_{i\in\{1,2\}} \mathrm{epay}_{[3]}^i(d_{1,3},d_{2,3})\ldot \mathbf{0}
\end{align*}

\paragraph{\textbf{Roles of the four communicating sessions for the broker}}
\begin{align*}
    R_\mathrm{broker}^\mathrm{auc} \overset{\mathrm{def}}{=} ~& \sum_{i\in \{1,2\}} (a_{1,i+1}?\mathrm{bid}\ldot\rec{X_1}(a_{1,4-i}!\mathrm{quote}\ldot(a_{1,i+1}!\mathrm{invoice}\ldot\mathbf{0}\ldot +\\
    &  \quad a_{1,4-i}?\mathrm{bid}\ldot a_{1,i+1}!\mathrm{quote}\ldot(a_{1,i+1}?\mathrm{bid}\ldot X_1+a_{1,4-i}!\mathrm{invoice}\ldot\mathbf{0})))) \\
    R_\mathrm{broker}^\mathrm{sTran} \overset{\mathrm{def}}{=} ~&  c_{2,3}!\mathrm{prepaid}\ldot c_{1,2}?\mathrm{confirm}\ldot c_{2,3}!\mathrm{payment}\ldot\mathbf{0}\\
    R_\mathrm{broker}^\mathrm{dTran} \overset{\mathrm{def}}{=} ~& b_{2,3}!\mathrm{price}\ldot\mathbf{0} \qquad
     R_\mathrm{broker}^\mathrm{epay} \overset{\mathrm{def}}{=} d_{2,3}?\mathrm{transfer}\ldot\mathbf{0}
\end{align*}

\paragraph{\textbf{Roles of the four communicating sessions for the buyers}} Let $j\in \{1,2\}$.
\begin{align*}\
    R^\mathrm{auc}_{\mathrm{buyer}_{j}} \overset{\mathrm{def}}{=} ~& a_{1,j+1}!\mathrm{bid}\ldot\rec{X_1}(a_{1,j+1}?\mathrm{quote}\ldot a_{1,j+1}!\mathrm{bid}\ldot X ~+ \\
    & \quad  a_{1,j+1}?\mathrm{invoice}\ldot \mathbf{0})+ a_{1,j+1}?\mathrm{quote}\ldot\\
     & \qquad \rec{X_2}(a_{1,j+1}!\mathrm{bid}\ldot (a_{1,j+1}?\mathrm{invoice}\ldot \mathbf{0}+a_{1,j+1}?\mathrm{quote}\ldot X_2)) \\
    R^\mathrm{dTran}_{\mathrm{buyer}_{j}} \overset{\mathrm{def}}{=} ~& b_{1,3}!\mathrm{payment}\ldot b_{1,3}?\mathrm{order}\ldot \mathbf{0} \\
    R^\mathrm{sTran}_{\mathrm{buyer}_{j}} \overset{\mathrm{def}}{=} ~& c_{1,3}?\mathrm{order}\ldot \mathrm{c}_{1,2}!\mathrm{confirm}\ldot \mathbf{0} \qquad
    R^\mathrm{epay}_{\mathrm{buyer}_{j}} \overset{\mathrm{def}}{=}   d_{1,3}!\mathrm{amount}\ldot \mathbf{0}
\end{align*}

\paragraph{\textbf{Roles of DTransaction and DTransaction for the seller}}
\begin{align*}
    R^{\mathrm{dTran}}_{\mathrm{seller}}\overset{\mathrm{def}}{=} ~& b_{2,3}?\mathrm{price}\ldot b_{1,3}?\mathrm{payment}\ldot b_{1,3}!\mathrm{order}\ldot \mathbf{0}\\
    R^{\mathrm{sTran}}_{\mathrm{seller}}\overset{\mathrm{def}}{=} ~& c_{2,3}?\mathrm{prepaid}\ldot c_{1,3}!\mathrm{order}\ldot c_{1,3}!\mathrm{payment}\ldot \mathbf{0}
\end{align*}

\paragraph{\textbf{The role of EPay for the bank}}
\begin{align*}
    R^\mathrm{epay}_{\mathrm{bank}} \overset{\mathrm{def}}{=}  d_{1,3}?\mathrm{amount}\ldot d_{2,3}!\mathrm{transfer}\ldot \mathbf{0}
\end{align*}

\section{Derivation of $\Gamma_{\mathrm{prt}}\vdash P_{\mathrm{broker}}\triangleright R_{\mathrm{broker}}^{\mathrm{all}}$}

This part of the appendix is dedicated to detailing a derivation of the type judgement $\Gamma_{\mathrm{prt}}\vdash P_{\mathrm{broker}}\triangleright R_{\mathrm{broker}}^{\mathrm{all}}$ in Proposition \ref{pr:typeexample}. Derivations of other type judgements in Proposition \ref{pr:typeexample} can be constructed in a similar way.
\begin{enumerate}
  \item by \textsc{[T-nil]}\ref{ty:t-tml}:
   $$ \Gamma_{\mathrm{prt}} \vdash  \mathbf{0}\triangleright \mathbf{0}\circ (c_{1,2},c_{1,3},c_{2,3}):\mathbf{0}$$
  \item by \ref{ty:t-sr}:
     $$\Gamma_{\mathrm{prt}} \vdash R_\mathrm{broker}^\mathrm{sTran} \triangleright \mathbf{0}\circ (c_{1,2},c_{1,3},c_{2,3}):  R_\mathrm{broker}^\mathrm{sTran}$$
  \item by \ref{ty:t-tml}\ref{ty:t-sr}:
  \begin{align*}
    & \Gamma_{\mathrm{prt}} \vdash  d_{1,3}?\mathrm{transfer}\ldot R_\mathrm{broker}^\mathrm{sTran} \triangleright \mathbf{0}\circ (d_{1,3},d_{2,3}):R_{\mathrm{broker}}^{\mathrm{epay}},  (c_{1,2},c_{1,3},c_{2,3}):  R_\mathrm{broker}^\mathrm{sTran}
  \end{align*}
  \item let $i\in \{1,2\}$ and \label{drv:8}
  \begin{align*}
    P_1^i \overset{\mathrm{def}}{=} \mathrm{sTran}_{[2]}^i(c_{1,2},c_{1,3},c_{2,3})\ldot \mathrm{epay}_{[2]}^i(d_{1,3},d_{2,3})\ldot d_{1,3}?\mathrm{transfer}\ldot  R_\mathrm{broker}^\mathrm{sTran}
  \end{align*}
  \item by \textsc{[T-ch]}\ref{ty:t-acc}:\label{drv:1}
  $$\Gamma_{\mathrm{prt}} \vdash P_1^i  \triangleright \mathrm{sTran}_{[2]}^i(c_{1,2},c_{1,3},c_{2,3})\ldot \mathrm{epay}_{[2]}^i(d_{1,3},d_{2,3})\ldot\mathbf{0}$$
  \item by \textsc{[T-nil]}\ref{ty:t-tml}\ref{ty:t-sr}\textsc{[T-ch]}\ref{ty:t-acc}: \label{drv:2}
  $$ \Gamma_{\mathrm{prt}} \vdash \mathrm{dTran}_{[2]}^i(b_{1,2},b_{2,3})\ldot b_{2,3}!\mathrm{price}\ldot\mathbf{0} ~\triangleright\mathrm{dTran}_{[2]}^i(b_{1,2},b_{2,3})\ldot \mathbf{0}$$
  \item let
  $$P_2^i=\mathrm{dTran}_{[2]}^i(b_{1,2},b_{2,3})\ldot \mathbf{0} + \mathrm{sTran}_{[2]}^i(c_{1,2},c_{1,3},c_{2,3})\ldot \mathrm{epay}_{[2]}^i(d_{1,3},d_{2,3})\ldot\mathbf{0}$$
  \item by \eqref{drv:1}\eqref{drv:2}\ref{ty:t-sum}:
  $$\Gamma_{\mathrm{prt}}\vdash P_{\mathrm{broker}}^i \triangleright P_2^i\qquad \Gamma_{\mathrm{prt}}\vdash P_{\mathrm{broker}}^{3-i} \triangleright P_2^{3-i}$$
  \item by \ref{ty:t-tml}\ref{ty:t-sr}:\label{drv:3}
  $$\Gamma_{\mathrm{prt}}\vdash a_{1,4-i}!\mathrm{invoice}\ldot P_{\mathrm{broker}}^{3-i} \triangleright P_2^{3-i}: (a_{1,2},a_{1,3}):a_{1,4-i}!\mathrm{invoice}\ldot\mathbf{0}$$
  \item by \ref{ty:t-var}\ref{ty:t-tml}\ref{ty:t-sr}: \label{drv:4}
  $$\Gamma_{\mathrm{prt}}\vdash a_{1,i+i}?\mathrm{bid}\ldot X \triangleright \mathbf{0}\circ  (a_{1,2},a_{1,3}):a_{1,i}?\mathrm{bid}\ldot X$$
  \item let
  $$P_3^i= a_{1,4-i}?\mathrm{bid}\ldot a_{1,i+1}!\mathrm{quote}\ldot
(a_{1,i+i}?\mathrm{bid}\ldot X+a_{1,4-i}!\mathrm{invoice}\ldot \mathbf{0})$$
  \item by \eqref{drv:3}\eqref{drv:4}\ref{ty:t-sum}\ref{ty:t-sr}\ref{ty:t-eq}: \label{drv:5}
   $$\Gamma_{\mathrm{prt}}\vdash P_3^i\{P_{\mathrm{broker}}^{3-i}/\mathbf{0}\}\triangleright P_2^{3-i} \circ (a_{1,2},a_{1,3}):P_3^i$$
  \item by \ref{ty:t-tml}\ref{ty:t-sr}: \label{drv:6}
  $$\Gamma_{\mathrm{prt}}\vdash a_{1,i+1}!\hbox{invoice}\ldot P_{\mathrm{broker}}^i\triangleright P_2^i\circ (a_{1,2},a_{1,3}): a_{1,i+1}!\hbox{invoice}\ldot\mathbf{0}$$
  \item let
  \begin{align*}
    P_4^i= \;& a_{1,4-i}!\mathrm{quote}\ldot (a_{1,i+1}!\hbox{invoice}\ldot P_{\mathrm{broker}}^i+ P_3^i\{P_{\mathrm{broker}}^{3-i}/\mathbf{0}\})\\
    P_5^i= \;& a_{1,4-i}!\mathrm{quote}\ldot (a_{1,i+1}!\hbox{invoice}\ldot \mathbf{0}+ P_3^i)
  \end{align*}
  \item by \eqref{drv:5}\eqref{drv:6}\ref{ty:t-sum}\ref{ty:t-sr}:
  $$\Gamma_{\mathrm{prt}}\vdash P_4^i \triangleright P_2^i+P_2^{3-i} \circ a_{1,4-i}!\hbox{quote}\ldot(a_{1,i+1}!\hbox{invoice}\ldot \mathbf{0} +P_3^i) $$
  \item by \ref{ty:t-rec}\ref{ty:t-sr}:\label{drv:7}
  $$\Gamma_{\mathrm{prt}}\vdash a_{1,i+i}?\mathrm{bid}\ldot \rec{X} P_4^i \triangleright P_2^i+P_2^{3-i}\circ a_{1,i+1}?\hbox{bid}\ldot \rec{X}P_5^i$$
  \item by \eqref{drv:7}\eqref{drv:8}\ref{ty:t-sum} ($P_2^1+P_2^{3-1}\equiv P_2^2+P_2^{3-2}$):
  $$\Gamma_{\mathrm{prt}}\vdash \sum_{i\in \{1,2\}}P_4^i\triangleright P_2^1+P_2^2\circ R_{\mathrm{broker}}^{\mathrm{auc}}$$
  \item by \ref{ty:t-inv}:
  $$\Gamma_{\mathrm{prt}}\vdash P_{\mathrm{broker}}\triangleright R_{\mathrm{broker}}^{\mathrm{all}}$$
\end{enumerate}

This finishes the derivation.

\section{More Examples}

We present two more examples to show the utility of our two-level session types in expressing the scenario-based specification of systems.
\paragraph{\textbf{Client-server system}} The first one is a client-server system, which consists of one client, two servers and a configurator. The client attempts to make requests to the servers and the configurator co-ordinates the client and two servers so that the client can only call the available server(s). Both servers have two states: they are either in the normal working order or preparing to update their data bases and shut down the service temporarily. The servers inform the configurator of their states in their conversations. Before the client calls the servers, the configurator tells them whether the servers are ready to take requests.

The following is a formulation of the session specification in our two-level session types, where $C$ is the client, $S_1, S_2$ are two servers, $F$ is the configurator, \hbem{CSsystem} is an integrating session type, and \hbem{Control}, \hbem{Initi} and \hbem{Service} are communicating session types.
\begin{align*}
    \hbem{CSsystem} \overset{\mathrm{def}}{=} & ~ \langle F,S_1, S_2: \hbem{Control}\rangle\{\} \\
    & ~\quad \otimes \mu \mathbf{t}.(\bigoplus_{i\in \{1,2\}}\langle C,F:\hbem{Initi}\rangle \{\};
     \langle C,S_i: \hbem{Service}\rangle \{ \};\mathbf{t}) \\
    \hbem{Control} \overset{\mathrm{def}}{=} & ~  \mu \mathbf{t}.(\bigoplus_{j\in \{2,3\}} \langle j,1:\hbox{update}\rangle \rightarrow \langle j,1:\hbox{ready}\rangle \rightarrow \mathbf{t}) \\
    \hbem{Initi} \overset{\mathrm{def}}{=} & ~  \bigoplus_{k\in \{1,2\}}\langle 1,2: \hbox{ping}_k\rangle \rightarrow (\langle 1,2:\hbox{yes}\rightarrow \mathtt{end} \rangle \oplus \langle 1,2:\hbox{no} \rangle\rightarrow \mathtt{end}) \\
    \hbem{Service} \overset{\mathrm{def}}{=} & ~ \langle 1,2 :\hbox{request}\rangle \rightarrow \langle 2,1:\hbox{return}\rangle \rightarrow \mathtt{end}
\end{align*}

The formulation captures the intuitive and coarse-grained understanding of the conversations between the four components. First, the conversations consists of three parts, represented by three communicating sessions. Second, the relationship of these sessions is described by \hbem{CSsystem}, revealing the most essential design decisions of the system. For example, \hbem{Service} happens after \hbem{Initi} and together they form a recursive session. \hbem{Control} is also recursive and proceeds independent of the other two communicating sessions. Of course, many design details are to be worked out in the later development stage. For example, if the configurator replies `no' to the client's pinging action in \hbem{Initi} , then the client is not allowed to initiate \hbem{Service}. Also, the messages received by the configurator in \hbem{Control} should affect its replies to the client's pinging action in \hbem{Init}.

\paragraph{\textbf{Quote request}} The second example is a quote request protocol which is modified and simplified from the one in \cite{ydbh10}. The protocol involves three agents, i.e.~a buyer, a supplier, and a manufacturer, and consists of two parts: the first part is a conversation between the buyer and the supplier, in which the price of some item or good is negotiated; the second part, which is nested within in the first part, is for the supplier to confirm the price with the manufacturer.

As before, the formulate consists of one integrating session and several (here is two) communicating sessions. $B$ stands the buyer, $S$ the supplier, and $M$ the manufacturer.
\begin{align*}
    \hbem{QuoteReq} \overset{\mathrm{def}}{=} & ~ \langle B,S:\hbem{Negotn} \rangle \{ \langle S,M:\hbem{Confirm}\rangle\{\}\} \\
    \hbem{Negotn} \overset{\mathrm{def}}{=} & ~ \langle 1,2:\hbox{item}\rangle \rightarrow \langle 2,1:\hbox{quote}\rangle \rightarrow (\langle 1,2:\hbox{accepted}\rangle\rightarrow \mathtt{end}~ \oplus \\
    & ~ \quad  \langle 1,2:\hbox{newquote}\rangle \rightarrow ( \langle 2,1:\hbox{accepted}\rangle \rightarrow \mathtt{end} ~\oplus  \\
    & ~ \qquad \langle 2,1:\hbox{rejected}\rangle \rightarrow \mathtt{end}))\\
    \hbem{Confirm} \overset{\mathrm{def}}{=} & ~ \langle 1,2:\hbox{quote}\rangle\rightarrow (\langle 1,2:\hbox{yes}\rightarrow \mathtt{end} \rangle \oplus \langle 1,2:\hbox{no} \rangle\rightarrow \mathtt{end})
\end{align*}

This example shows the necessity to distinguish protocol calling and protocol nesting (c.f.~discussions in Sect.~\ref{sec:relatedwork}). Because, as far as the protocol is concerned, it suffices to indicate the nesting relationship between \hbem{Negotn} and \hbem{Confirm}. Without specifying the nesting position of \hbem{Confirm} in \hbem{Negotn}, \hbem{Negotn} describes a complete conversation between the buyer and the supplier.

\section{Proof Details}
\subsubsection*{Proof of Theorem \ref{th:typeinf}}
\begin{proof}
The proof of is a standard proof of decidability of type inference. Because of \ref{ty:t-eq}, a process has infinite many typing, but we show that we can compute a `principal' typing for each process such that the process has a `principal' typing if and only if it is typable.

First, for each $P$, we compute a set $\mathrm{sub}_{\2}(P)$ (resp.~$\mathrm{sub}_{\sqcup}(P)$) which is the smallest set such that
\begin{enumerate}
  \item if $\langle P_1, P_2\rangle \in \mathrm{sub}_{\2}(P)$ (resp.~$\mathrm{sub}_{\sqcup}(P)$) then $ P_1\2 P_2 \equiv P$ (resp.~$P_1\sqcup P_2\equiv P$), and
  \item if $ Q_1\2 Q_2 \equiv P$ (resp.~$Q_1\sqcup Q_2\equiv P$) then there are $ P_1, P_2$ such that $P_1\equiv Q_1$, $P_2\equiv Q_2$ and $\langle P_1, P_2\rangle \in \mathrm{sub}_{\2}(P)$ (resp.~$\mathrm{sub}_{\sqcup}(P)$).
\end{enumerate}

The two sets are decidable because $\equiv$ is decidable (Lemma \ref{le:process-equivalence}). $\mathrm{sub}_{\2}(\Delta)$ is defined as follows: $\langle \Delta_1,\Delta_2\rangle \in \mathrm{sub}_{\2}(\Delta)$ if and only if $|\Delta_1|=|\Delta_2|=|\Delta|$ and, for each $1\leq i\leq |\Delta|$,  $\langle P_1^i , P_2^i\rangle \in \mathrm{sub}_{\2}(P^i)$ where $\Delta_1[i]= \tilde{c}: P_1^i$ , $ \Delta_2[i]= \tilde{c}:P_2^i$ and $P^i=\tilde{c}:\Delta[i]$ for some $\tilde{c}$. $\mathrm{sub}_{\sqcup}(\Delta)$ is defined similarly. Note that if $\langle \Delta_1,\Delta_2\rangle \in \mathrm{sub}_{\2}(\Delta)$ (resp.~$\mathrm{sub}_{\sqcup}(\Delta)$), then $\Delta_1\asymp \Delta_2$.

A \emph{principal} typing of $P$ under $\Gamma$ is a typing derived by the rules in Figures \ref{ty:rule} except \ref{ty:t-eq}, and plus the following two rules:
\begin{gather*}
\dfrac{\Gamma \vdash P\triangleright R\circ \Delta \quad \Gamma \vdash P'\triangleright R'\circ \Delta'\quad \langle  R, R'\rangle\in \mathrm{sub}_{\2}(R'')\quad \langle \Delta,\Delta'\rangle\in \mathrm{set}_{\2}(\Delta'')}{\Gamma\vdash P\2 P' \triangleright R''\circ \Delta''} \tag*{\,[\textsc{T-com+}]}\label{ty:t-com+}\\
\dfrac{\Gamma\vdash P\triangleright R\circ \Delta \quad \Gamma\vdash P'\triangleright R'\circ \Delta'\quad \langle  R, R'\rangle\in \mathrm{sub}_{\tsum}(R'')\quad \langle \Delta,\Delta'\rangle\in \mathrm{set}_{\tsum}(\Delta'')}{\Gamma\vdash P+ P' \triangleright R''\circ \Delta''}  \tag*{\,[\textsc{T-sum+}]}\label{ty:t-sum+}\end{gather*}

We have the following lemma:
\begin{le1}\label{le:principaltype} $\Gamma \vdash P\triangleright R\circ \Delta$ if and only if $P$ has a principal typing under $\Gamma$.
\end{le1}

The right-to-left direction of the lemma is obvious. For the other direction, we suppose $\Gamma \vdash P\triangleright R\circ \Delta$. In the derivative procedure, if commutative and associative laws for $\2$ and $+$ are applied in \ref{ty:t-eq}, we have the same derivation by the additional two derived rules, and whenever other structural laws in Figure \ref{cong} are applied in \ref{ty:t-eq}, we just omit them. In this manner, we will obtain a principal typing for $P$. Therefore, the type inference of the type system is decidable.\end{proof}

Note that if $R'\circ\Delta'$, say, is the principal typing of $P$, then by Theorem \ref{th:typcon} (to be proved) $R\equiv R'$ and $\lrceil{\Delta}=\lrceil{\Delta'}$.

\subsubsection*{Proof of Theorem \ref{th:subcon}}

\begin{proof}Suppose $\Gamma\vdash P$ and $P\equiv P'$.
The proof is by induction on the derivation of $P\equiv P'$. The proof is divided into two parts. First, we show that the each rule in Figure \ref{cong} and its symmetric form respect the above theorem. Here we detail one of the most tricky rules:
\begin{gather*}
    (\nu a)P_1 \2 P_2 \equiv (\nu a )(P_1\2 P_2) \hbox{ if } a\notin\mathrm{fc}(P_2)
\end{gather*}
(1) We first suppose $P=(\nu a)P_1\2 P_2$, $\Gamma\vdash P \triangleright R\circ \Delta$ and $a\notin\mathrm{fc}(P_2)$. Because $\Gamma\vdash P \triangleright R\circ \Delta$ is derived by \ref{ty:t-com} and possibly by \ref{ty:t-eq}, \ref{ty:t-tml} and/or \ref{ty:t-tmr} (for one or more times), it can be verified that $\Gamma\vdash (\nu a)P_1\triangleright R_1\circ \Delta_1$, $\Gamma\vdash P_2\triangleright R_2\circ \Delta_2$, $\Delta_1\asymp \Delta_2$, $R\equiv R_1\2 R_2$, and $\lrceil{\Delta}\equiv\lrceil{\Delta_1\2 \Delta_2}$ for some $R_1,R_2,\Delta_1,\Delta_2$. Here we have two possibilities. (1.1) Suppose $a\in \mathrm{ch}(\Delta_1)$. Thus, $\Gamma\vdash (\nu a)P_1\triangleright R_1\circ \Delta_1$ is derived by \ref{ty:t-hid} and possibly by \ref{ty:t-eq}\ref{ty:t-tml}\ref{ty:t-tmr}, and, we have that $\Gamma\vdash P_1\triangleright R_1\circ \Delta_1',\tilde{c}:Q,\Delta_1''$, $a\in \tilde{c}$, and $\Delta_1\equiv\tilde{c}_1:\mathbf{0},\ldots,\tilde{c}_m:\mathbf{0},\Delta_1',\tilde{c}\backslash a:Q,\Delta_1'', \tilde{c}_{m+1}:\mathbf{0}, \ldots, \tilde{c}_{m+n}:\mathbf{0}$.
No matter $a\in\bigcup_{i=1}^{m+n} \tilde{c}_i$ or not, we can rewrite the processes of type derivations for $P_1$ and $P_2$ to obtain type judgements $\Gamma\vdash P_1\triangleright R_1\circ \Delta_3$ and $\Gamma\vdash P_2\triangleright R_2\circ \Delta_2$ such that $\Delta_3\asymp \Delta_4$ and $\lrceil{\Delta_3\2\Delta_4}=\lrceil{\Delta_1\2\Delta_2}$ (when applying \ref{ty:t-tml} or \ref{ty:t-tmr} to prefix $ \tilde{b}$ for some $\tilde{b}$, we prefix $ \tilde{b}/a$ or some $\tilde{b}'$ such that $\tilde{b'}\backslash a=\tilde{b}$ instead). Note that the rewritten derivations are based on $a\in \mathrm{fc}(P_2)$ and the channel assumption (cf.~Sect.~\ref{sec:calculus}). Then, we apply \ref{ty:t-hid} to type $(\nu a )(P_1\2 P_2)$ and obtain the desired result. (1.2) Suppose $a\not\in \mathrm{ch}(\Delta_1)$. Thus, $\Gamma\vdash (\nu a)P_1\triangleright R_1\circ \Delta_1$ is derived by \ref{ty:t-vei} and possibly by \ref{ty:t-eq}\ref{ty:t-tml}\ref{ty:t-tmr}, and we have that $\Gamma\vdash P_1\triangleright R_1\circ \Delta_1'$ and $\Delta_1\equiv\tilde{c}_1:\mathbf{0},\ldots,\tilde{c}_m:\mathbf{0},\Delta_1', \tilde{c}_{m+1}:\mathbf{0}, \ldots, \tilde{c}_{m+n}:\mathbf{0}$. Similarly, we rewrite the type derivation for $P_2$ and obtain $\Gamma\vdash P_2\triangleright R_2\circ \Delta_2'$ such that $\Delta_1\asymp \Delta_2'$ and $\lrceil{\Delta_2'}\equiv \lrceil{\Delta_2}$. Then, apply \ref{ty:t-vei} to $\Gamma\vdash P_2\triangleright R_2\circ \Delta_2'$ and obtain the desired result. (2) Then, we suppose $P=(\nu a)(P_1\2 P_2)$, $\Gamma\vdash P \triangleright R\circ \Delta$ and $a\notin\mathrm{fc}(P_2)$. The treatment is similar to the first case.

The second part of the proof is to show that the laws of congruence respect the theorem. We choose to deal with the following rule:
\begin{gather*}
    \dfrac{P_1\equiv P_2}{\rec{X}P_1\equiv \rec{X}P_2}
\end{gather*}
We suppose $P=\rec{X}P_1$, $P'=\rec{X} P_2$, $P_1\equiv P_2$, and $\Gamma \vdash P\triangleright R\circ \Delta$ where $\Delta\equiv \tilde{c}_1:Q_1,\ldots, \tilde{c}_n:Q_n$. Because $\Gamma \vdash P\triangleright R\circ \Delta$ is derived by \ref{ty:t-rec} (and \ref{ty:t-eq}\ref{ty:t-tml}\ref{ty:t-tmr}, possibly), we have $\Gamma, X\vdash P_1\triangleright R_1\circ \tilde{c}_1:Q_1',\ldots, \tilde{c}_n:Q_n'$ such that $R_1\equiv R^{[X]}$ and $Q_i'\equiv Q_i^{[X]}$ for each $1\leq i\leq n$. Since $P_1\equiv P_2$, by induction hypotheses, $\Gamma, X\vdash P_2\triangleright R_2\circ \tilde{c}_1':Q_1'', \ldots, \tilde{c}_n':Q_n''$ for some $R_2$, $\tilde{c}_i'$ and $Q_i''$ for each $1\leq i\leq n$ such that $R_2\equiv R_1$ and $Q_1'\2 \ldots \2 Q_n'\equiv Q_1''\2 \ldots Q_n''$. By \ref{ty:t-rec}, we have that $\Gamma \vdash \rec{X}P_2\triangleright R \circ \Delta'$ where $R\equiv R_2^{[X]}$ and $\Delta'\equiv\tilde{c}_1': {Q_1''}^{[X]},\ldots \tilde{c}_n':{Q_n''}^{[X]}$. Therefore, we have that $R\equiv R'$ and $\lrceil{\Delta}\equiv\lrceil{\Delta'}$.
\end{proof}

\subsubsection*{Proof of Theorem \ref{th:subred}}

\begin{proof} The proof is by induction on the derivation of $P\overset{\alpha}{\longrightarrow} P'$ according to rules in Figure \ref{tran} and depends on the value of $\alpha$. The following only covers the most typical cases.

(1) Suppose $P=\alpha. P'$. In this case, $\alpha$ has three possible forms: $a\S v$, $\bar{a}_{[2..n]}(\tilde{c})$ or $a_{[k]}(\tilde{c})$ where $\S\in \{?,!\}$. First, we let $\alpha=a\S v$. Because $\Gamma\vdash P\triangleright R\circ \Delta$ is derived by \ref{ty:t-sr} and possibly by \ref{ty:t-eq}\ref{ty:t-tml}\ref{ty:t-tmr} (for one or more times), we have that $\Gamma\vdash P'\triangleright R'\circ \Delta'$ for some $R',\Delta'$ such that $R\equiv R'$, $\lceil \Delta'\rceil \equiv \lceil \tilde{c}:Q, \Delta''\rceil$, and $\lceil\Delta\rceil\equiv \lceil \tilde{c}:a\S v.Q, \Delta''\rceil$ for some $\tilde{c}, Q,\Delta''$. We have that $\lceil \Delta\rceil \overset{\alpha}{\longrightarrow}_\succ \lceil\Delta'\rceil$. Then, let $\alpha=\bar{a}_{[2..n]}(\tilde{c})$. Because $\Gamma\vdash P\triangleright R\circ \Delta$ is derived by \ref{ty:t-inv} (and possibly \ref{ty:t-eq}\ref{ty:t-tml}\ref{ty:t-tmr}), we have that $\Gamma\vdash P'\triangleright R'\circ \Delta'$, $R \equiv \bar{a}_{[2..n]}(\tilde{c}).R'$ (thus $R' \overset{\alpha}{\longrightarrow} R$) and $\lceil\Delta\rceil\equiv \lceil \Delta'\rceil\2 B{\upharpoonright}1\langle \tilde{c}\rangle $ where $\Gamma \vdash a\vdash B$. Lastly, let $\alpha=a_{[k]}(\tilde{c})$. Because $\Gamma\vdash P\triangleright R\circ \Delta$ is derived by \ref{ty:t-acc} (and possibly \ref{ty:t-eq}\ref{ty:t-tml}\ref{ty:t-tmr}), we have that $\Gamma\vdash P'\triangleright R'\circ \Delta'$, $R \equiv a_{[k]}(\tilde{c}).R'$ (thus $R' \overset{\alpha}{\longrightarrow} R$), and $\lceil\Delta\rceil\equiv \lceil \Delta'\rceil\2 B{\upharpoonright}1\langle \tilde{c}\rangle$ where $\Gamma \vdash a\vdash B$ and $2\leq k\in \mathrm{pid}(B)$.

(2) Let $\alpha=a\S v$ and suppose $P=P_1+P_2$ and $P\overset{\alpha}{\longrightarrow}P'$ is derived from $P_1\overset{\alpha}{\longrightarrow}P'$ (the treatment is similar of it derived from $P_2\overset{\alpha}{\longrightarrow}P'$). By \ref{ty:t-sum} (and possibly by \ref{ty:t-eq}\ref{ty:t-tml}\ref{ty:t-tmr}), we have that $\Gamma\vdash P_1\triangleright R_1\circ \Delta_1$, $\Gamma\vdash P_2\triangleright R_2\circ \Delta_2$, $R\equiv R_1\tsum R_2$, and $\lceil \Delta\rceil\equiv\lceil\Delta_1\tsum \Delta_2\rceil$. By induction hypotheses, $\Gamma\vdash P'\triangleright R'\circ \Delta'$, $ R_1\succ R'$ and $\lceil \Delta_1\rceil \overset{\alpha}{\longrightarrow}_\succ \lceil \Delta'\rceil$. Hence, $ R\succ R'$ and $\lceil \Delta\rceil \overset{\alpha}{\longrightarrow}_\succ \lceil\Delta'\rceil$.

(3) Let $\alpha=\tau$ and $P=P_1\2 P_2$. Here we have two subcases. (3.1) Suppose $P\overset{\tau}{\longrightarrow}P'$ is derived from $P_1\overset{\tau}{\longrightarrow}P'$ (or $P_2\overset{\alpha}{\longrightarrow}P'$). The treatment for this subcase is relatively simple and similar to the last case and thus we omit it. (3.2) $P\overset{\tau}{\longrightarrow}P'$ is derived from $P_1\overset{a!v}{\longrightarrow}P_1'$ and $P_2\overset{a?v}{\longrightarrow}P_2'$ and $P'=P_1'\2 P_2'$. Because $\Gamma\vdash P \triangleright R\circ \Delta$ is derived by \ref{ty:t-com} (and possibly \ref{ty:t-eq}\ref{ty:t-tml}\ref{ty:t-tmr}), we have that  $\Gamma\vdash P_1\triangleright R_1\circ \Delta_1$ and $\Gamma\vdash P_2\triangleright R_2\circ \Delta_2$ for some $R_1,R_2,\Delta_1,\Delta_2$ such that $ R\equiv R_1\2 R_2$, and $\lceil \Delta\rceil \equiv \lceil \Delta_1\2 \Delta_2\rceil$. By induction hypotheses, $\Gamma\vdash P_1'\triangleright R_1'\circ \Delta_1'$ and $\Gamma\vdash P_2'\triangleright R_2'\circ \Delta_2'$ for some $R_1',R_2', \Delta_1',\Delta_2'$ such that $R_1\succ R_1'$, $R_2\succ  R_2'$, $\lceil \Delta_1\rceil \overset{a!v}{\longrightarrow}_\succ \lceil \Delta_1'\rceil$ and $\lceil \Delta_2\rceil \overset{a?v}{\longrightarrow}_\succ \lceil \Delta_2'\rceil$. Also, $\Delta_1'\asymp\Delta_2'$. Hence, $\Gamma\vdash P'\triangleright R_1'\2 R_2'\circ \Delta_1'\2 \Delta_2'$, $R_1\2 R_2\succ R_1'\2 R_2'$ and $\lrceil{\Delta_1\2 \Delta_2}\overset{\tau}{\longrightarrow}_\succ \lrceil{\Delta_1'\2 \Delta_2'}$. (3.3) $P\overset{\tau}{\longrightarrow}P'$ is derived from $P_1\overset{\bar{a}_{[2..n]}(\tilde{c})}{\longrightarrow}P_2'$ and $P_i\overset{a_{[i]}(\tilde{c})}{\longrightarrow}P_i'$ for each $2\leq i\leq n$. Suppose $\Gamma\vdash a\triangle B$. Because $\Gamma\vdash P \triangleright R\circ \Delta$ is derived by \ref{ty:t-com} for $n-1$ times (and possibly \ref{ty:t-eq}\ref{ty:t-tml}\ref{ty:t-tmr}), we have that $\Gamma\vdash P_1\triangleright R_1\circ \Delta_1$ and $\Gamma\vdash P_i\triangle R_i\circ\Delta_i$ ($2\leq i\leq n$) for some $R_1,R_i,\Delta_1,\Delta_i$ such that $R\equiv R_1\2 R_2\2 \ldots\2 R_n$ and $\lrceil{\Delta}\equiv \lrceil{\Delta_1}\2 \lrceil{\Delta_2}\2\ldots\2 \lrceil{\Delta_n}$. By induction hypotheses, $\Gamma\vdash P_1'\triangleright R_1'\circ \Delta_1'$ and $\Gamma\vdash P_i'\triangle R_i'\circ\Delta_i'$ ($2\leq i\leq n$) for some $R_1',R_i',\Delta_1',\Delta_i'$ such that $R_1\overset{\bar{a}_{[2..n]}(\tilde{c})}{\longrightarrow} R_1'$, $R_i\overset{a_{[i]}(\tilde{c})}{\longrightarrow} R_i'$, $\lrceil{\Delta_1'}\equiv \lrceil{\Delta_1}\2 B{\upharpoonright}1\langle\tilde{c}\rangle $, and $\lrceil{\Delta_i'}\equiv  \lrceil{\Delta_1}\2 B{\upharpoonright}i\langle\tilde{c}\rangle $ for each $2\leq i\leq n$. Also, $\Delta_1'\asymp\Delta_2'\asymp\ldots \asymp\Delta_n'$. Therefore, $\lrceil{\Delta_1'\2\ldots\2 \Delta_n'}\equiv B{\upharpoonright}1\langle\tilde{c}\rangle\2\ldots\2 B{\upharpoonright}n\langle\tilde{c}\rangle\2\lrceil{\Delta_1\2\ldots \2\Delta_n}$.

(4) Let $\alpha=a\S v$. Suppose $Q\equiv P$, $Q'\equiv P'$, and $P\overset{\alpha}{\longrightarrow}P'$ is derived from $Q\overset{\alpha}{\longrightarrow}Q'$. By Theorem \ref{th:subcon}, $\Gamma\vdash Q\triangleright R_1\circ \Delta_1$ such that $R_1\equiv R$ and $\lrceil{\Delta_1}\equiv \lrceil{\Delta}$. By induction hypotheses, $\Gamma\vdash Q'\triangleright R_1'\circ \Delta_1'$ such that $R_1\succ R_1'$ and $\lrceil{\Delta_1}\overset{\alpha}{\longrightarrow}_\succ \lrceil{\Delta_1'}$. Then, by Theorem \ref{th:subcon} again, we have the desired result.
\end{proof}

\subsubsection*{Proof of Theorem \ref{th:typcon}}

\begin{proof}The proof is by induction on the derivation of $\Gamma\vdash P\triangleright R\circ \Delta$ and $\Gamma\vdash P $ $\triangleright\;R'\circ \Delta'$ according to rules in Figure \ref{ty:rule}. We detail two cases.
(1) Suppose $P=\bar{a}_{[2..n]}(\tilde{c}).P_1$ and $\Gamma\vdash P\triangleright R\circ \Delta$ is derived by \ref{ty:t-inv}. Let $\Gamma\vdash a\triangleright B$ and $|\mathrm{pid}(B)|=n$. Thus, $\Gamma \vdash P_1 \triangleright R_1\circ \Delta_1 $ for some $R_1,\Delta_1$ such that $R=\bar{a}_{[2..n]}(\tilde{c}).R_1$ and $\Delta_1= \tilde{c}: B{\upharpoonright} 1,\Delta$. Also, $\Gamma\vdash P\triangleright R'\circ \Delta'$ and possibly \ref{ty:t-eq}\ref{ty:t-tml}\ref{ty:t-tmr} for one or more times. Thus, $\Gamma \vdash P_1 \triangleright R_1'\circ \Delta_1' $ for some $R_1',\Delta_1'$ such that $R'\equiv \bar{a}_{[2..n]}(\tilde{c}).R_1'$ and $\lrceil{\Delta_1'}\equiv B{\upharpoonright}1\2 \lrceil{\Delta'}$. By induction hypotheses, $R_1\equiv R_1'$ and $\lrceil{\Delta_1}=\lrceil{\Delta_1'}$. Therefore, $R\equiv R'$ and $\lrceil{\Delta}=\lrceil{\Delta'}$.
(2) Suppose $P=\rec{X} P'$, and $\Gamma\vdash P\triangleright R\circ \Delta$ and $\Gamma\vdash P\triangleright R'\circ \Delta'$ are derived by \ref{ty:t-rec}. Let $\Gamma\vdash P'\triangleright R_1\circ \tilde{c}_1:Q_1^1,\ldots,\tilde{c}_n:Q_n^1$ where $R_1^{[X]}=R$ and $\tilde{c}_1:{Q_1^1}^{[X]},\ldots,\tilde{c}_n:{Q_1^n}^{[X]}=\Delta$, and $\Gamma\vdash P'\triangleright R_2\circ \tilde{c}_1:Q_1^2,\ldots,\tilde{c}_n:Q_n^2$ where $R_2^{[X]}=R'$ and $\tilde{c}_1:{Q_2^1}^{[X]},\ldots,\tilde{c}_n:{Q_2^n}^{[X]}=\Delta'$. By induction hypotheses, $R_1\equiv R_2$ and $\lrceil{Q_1^1\2\ldots \2 Q_n^1}\equiv \lrceil{Q_1^2\2\ldots \2 Q_n^2}$ for each $1\leq i\leq n$. Thus, we have $R\equiv R'$ and $\lrceil{\Delta}\equiv \lrceil{\Delta'}$.
\end{proof}

\subsubsection*{Proof of of Theorem \ref{th:chanpri}}

\begin{proof}(Sketch) This lemma is guaranteed by the projection of sessions into roles and the generation of fresh channels in the session establishment. A formal proof is by induction on $\mathrm{Sys}\overset{\tau}{\longrightarrow}_\ast (\nu \tilde{a})(1:P_1\2 \ldots \2 n:P_n)$.
\end{proof}

\subsubsection*{Proof of Theorem \ref{th:sesfid}}

\begin{proof}We first put forward two lemmas, whose proofs are by the syntax of $B$ or $A$.

\begin{le1}\label{le:basictran}
(1) If $B\overset{p,q:v}{\longrightarrow}B'$ then $B{\upharpoonright}p\langle \tilde{c}\rangle \overset{b?v}{\longrightarrow}B'{\upharpoonright}p\langle \tilde{c}'\rangle $ and $B{\upharpoonright}q\langle \tilde{c}\rangle \overset{b!v}{\longrightarrow}B'{\upharpoonright}q\langle \tilde{c}'\rangle $ for some $b\in \tilde{c}\supseteq \tilde{c}'$. (2) If $B{\upharpoonright}p\langle \tilde{c}\rangle \overset{b?v}{\longrightarrow}P$ and $B{\upharpoonright}q\langle \tilde{c}\rangle \overset{b!v}{\longrightarrow}Q$ then there are $B',\tilde{c}'$ such that $B\overset{p,q:v}{\longrightarrow}B'$, $P\equiv B'{\upharpoonright}p\langle \tilde{c}'\rangle $, $Q\equiv B'{\upharpoonright}q\langle \tilde{c}'\rangle $ and $\tilde{c}'\subseteq \tilde{c}$.
\end{le1}

\begin{le1}\label{le:coordtran} Let $\tilde{p}=p_1,\ldots,p_m$ and $b$ marks $B$ in $A$.
(1) If $A\overset{\tilde{p}:B}{\longrightarrow}A'\otimes B\langle \tilde{p}\rangle$ then $A{\upharpoonright}p_1\overset{\bar{b}_{[2..m]}(\tilde{c})}{\longrightarrow} A'{\upharpoonright}p_1$ and $A{\upharpoonright}p_i\overset{b_{[i]}(\tilde{c})}{\longrightarrow} A'{\upharpoonright}p_i$ ($2\leq i\leq m$).
(2) If $A{\upharpoonright}p_1\overset{\bar{b}_{[2..m]}(\tilde{c})}{\longrightarrow} P_1$ and $A{\upharpoonright}p_i\overset{b_{[i]}(\tilde{c})}{\longrightarrow} P_i$ ($2\leq i\leq m$) then there is $A'$ such that $P_j \equiv A'{\upharpoonright}p_j$ ($1\leq j\leq m$) and $A\overset{\tilde{p}:B}{\longrightarrow}A'\otimes B\langle \tilde{p}\rangle$.
\end{le1}

Suppose $\mathrm{Sys}$ is well-typed by $A_\mathrm{spc}$. We construct an $\mathcal{R}$ such that $\langle P, S\rangle \in \mathcal{R}$ if and only if
\begin{itemize}
  \item $\mathrm{Sys}\overset{\tau}{\longrightarrow}_\ast P=(\nu \tilde{b})(1:P_1\2\ldots \2 n:P_n)$,
  \item $A_\mathrm{spc}\overset{\tau}{\longrightarrow}_\ast S=C\otimes B_1\otimes\ldots\otimes B_k$,
  \item for each $1\leq i\leq n$, $\Gamma \vdash P_i\triangleright R_i\circ \tilde{c}_1:Q^i_1,\ldots, \tilde{c}_k:Q^i_k$ where
  \begin{itemize}
    \item $C{\upharpoonright}i\langle \tilde{a}\rangle\succ R_i $,
    \item $B_j {\upharpoonright}i\langle \tilde{c_j}\rangle\succ Q^i_j$ for each $1\leq j\leq k$.
  \end{itemize}
\end{itemize}

First, we have that $\langle \mathrm{Sys},A_\mathrm{spc}\rangle \in \mathcal{R}$. Then, let $\langle P,S\rangle \in \mathcal{R}$ and suppose the above five induction hypotheses. Without loss of generality, we suppose (1) $P_1\overset{b!v}{\longrightarrow}P_1'$ and $P_2\overset{b?v}{\longrightarrow}P_2'$ or (2) $P_1\overset{\bar{b}_{[2..m]}(\tilde{c}_0)}{\longrightarrow}P_1'$ and $P_i\overset{b_{[i]}(\tilde{c}_0)}{\longrightarrow}P_i'$ ($2\leq i\leq m$) and $b$ marks $B_0$.

(1) Suppose $b\in \tilde{c}_j$. By Theorem \ref{th:subred}, $Q_j^1\overset{b!v}{\longrightarrow} {Q_j^1}'$ and $Q_j^2\overset{b?v}{\longrightarrow} {Q_j^2}'$. Thus, by Lemma \ref{le:basictran}, $B_j {\upharpoonright}1\langle \tilde{c}_j\rangle \overset{b!v}{\longrightarrow} B_j'{\upharpoonright}1\langle \tilde{c}_j'\rangle$ and $B_j {\upharpoonright}2\langle \tilde{c}_j\rangle \overset{b?v}{\longrightarrow} B_j'{\upharpoonright}2\langle \tilde{c}_j'\rangle$ for some $B_j',\tilde{c}'$ such that $B_j\overset{1,2:v}{\longrightarrow} B_j'$, $B_j'{\upharpoonright}1\langle \tilde{c}_j'\rangle\succ {Q_j^1}'$ and $B_j'{\upharpoonright}2\langle \tilde{c}_j'\rangle\succ {Q_j^2}'$. Also, $ B_j{\upharpoonright}i\langle \tilde{c}_j\rangle=B_j'{\upharpoonright}i\langle \tilde{c}_j'\rangle$ for each $3\leq i\leq n$.
Therefore, let $S'=C\otimes B_1'\otimes B_2 \ldots\otimes B_k$ and $P'=(\nu \tilde{b})(1:P_1'\2 2:P_2'\2 3:P_3\2 \ldots \2 P_n)$. We have that $\langle P',S'\rangle \in \mathcal{R}$. (2) By Theorem \ref{th:subred}, $R_1\overset{\bar{b}_{[2..m]}(\tilde{c}_0)}{\longrightarrow}R_1'$ and $R_i\overset{b_{[i]}(\tilde{c}_0)}{\longrightarrow}R_i'$ ($2\leq i\leq m$). Thus, by Lemma \ref{le:coordtran}, $C{\upharpoonright}\langle \tilde{a}\rangle \overset{\bar{b}_{[2..m]}(\tilde{c}_0)}{\longrightarrow} C'{\upharpoonright}\langle \tilde{a}\rangle $ and $C{\upharpoonright}\langle \tilde{a}\rangle \overset{b_{[i]}(\tilde{c}_0)}{\longrightarrow} C'{\upharpoonright}\langle \tilde{a}\rangle $ ($2\leq i\leq m$) for some $C',\tilde{a},\tilde{a}'$ such that $b\in\tilde{a}\supseteq \tilde{a}'$, $C\overset{1,\ldots,m:B_0}{\longrightarrow}C'$, $C'{\upharpoonright}l\langle \tilde{a}'\rangle\succ R_l'$ ($2\leq l\leq m$). Also, $C{\upharpoonright}l\langle \tilde{a}\rangle \equiv C'{\upharpoonright}l\langle \tilde{a}'\rangle $ for each $m+1\leq l\leq n$. Therefore, let $S'=C'\otimes B_0\otimes B_1\otimes \ldots\otimes B_k$ and $P'=(\nu \tilde{b})(1:P_1'\2 \ldots \2 m:P_m'\2 P_{m+1}\ldots \2 P_n)$. We have that $\langle P',S'\rangle \in \mathcal{R}$.
\end{proof}

\subsubsection{Proof of Theorem \ref{th:slicing}}

\begin{proof}
We prove a more general proposition: if
$\Gamma\vdash P\triangleright R\circ \Delta$ then
\begin{itemize}
  \item $P^{\chi_M}\equiv R$,
  \item if $\bar{a}_{[2..n]}(\tilde{c})\in \mathrm{act}(P)$ and $\Gamma\vdash a\triangleright B$, then $B{\upharpoonright}1\equiv P^{\chi_{\tilde{c}}}$,
  \item if $a_{[k]}(\tilde{c})\in \mathrm{act}(P)$ and $\Gamma\vdash a\triangleright B$, then $B{\upharpoonright}k\equiv P^{\chi_{\tilde{c}}}$,
  \item if $\Delta[i]=\tilde{c}:Q$, then $Q\equiv P^{\chi_{\tilde{c}}}$.
\end{itemize}

We observe that if the above proposition holds then Theorem \ref{th:slicing} immediately follows.
The proof of the proposition is by induction on the derivation of $\Gamma\vdash P\triangleright R\circ \Delta$ according to Figure \ref{ty:rule}. The basic cases are simple. For the non-basic cases, we choose to deal with two typical cases. (1) $P=a_{[2..n]}(\tilde{c}).P'$ and $\Gamma\vdash P\triangleright R\circ \Delta$ is derived by \ref{ty:t-inv}. Let $R=\bar{a}_{[2..n]}(\tilde{c}).R'$ and $\Delta=\tilde{c}:B{\upharpoonright}1,\Delta'$. By induction hypotheses, ${P'}^{\chi_M}\equiv R'$ and, thus, $P^{\chi_M}\equiv R$. Suppose $\bar{b}_{[2..m]}(\tilde{c}')\in \mathrm{act}(P)$, $\Gamma\vdash b\triangleright B$, and $\tilde{c}'\cap \mathrm{ch}(\Delta)=\varnothing$. If $b= a$ (and thus $\tilde{c}'=\tilde{c}$, then by induction hypotheses and the rule \ref{ty:t-inv}, ${P'}^{\chi_{\tilde{c}}}=B{\upharpoonright}1$. If $b\neq a$ $\ldots$, by induction hypotheses, we have the same result. The case of $b_{[l]}(\tilde{c}')$ is similar. Suppose $\Delta[i]=\tilde{c}':Q$. Then, $\tilde{c}'\cap \tilde{c}=\varnothing$ and $\Delta'[i+1]=\tilde{c}:Q$. Thus, $P^{\chi_{\tilde{c}'}}={P'}^{\chi_{\tilde{c}'}}$ and by induction hypotheses $Q\equiv P^{\chi_{\tilde{c}'}}$.
(2) $P=P_1\2 P_2$ and $\Gamma\vdash P\triangleright R\circ \Delta$ is derived by \ref{ty:t-com}. Let $R=R_1\2 R_2$, $\Delta=\Delta_1\2 \Delta_2$ and $\Gamma\vdash P_i\triangleright R_i\circ \Delta_i$ where $i\in \{1,2\}$. By induction hypotheses and the rule \ref{ty:t-com}, we can obtain the four propositions above. (N.B.~we have suppose $P$ does not contain the hiding operator, so the rules \ref{ty:t-hid} and \ref{ty:t-vei} are not applicable.)
\end{proof}


\end{document}